\documentclass[twocolumn]{aastex61}

\accepted{December 12, 2017}
\submitjournal{ApJ}

\shortauthors{Sokal et al.}
\shorttitle{Characterizing TW Hydra}

\usepackage{url}

\begin{document}


\title{Characterizing TW Hydra }


\author{Kimberly R. Sokal}
\affiliation{Department of Astronomy, The University of Texas at Austin, Austin, TX, 78712, USA}
\email{ksokal@utexas.edu}

\author{Casey P. Deen}
\affil{Max Planck Institut f{\"u}r Extraterrestrische Physik,  Giessenbachstrasse 1, Garching bei M{\"u}nchen, Germany}

\author{Gregory N. Mace}
\affiliation{Department of Astronomy, The University of Texas at Austin, Austin, TX, 78712, USA}

\author{Jae-Joon Lee}
\affil{Korea Astronomy and Space Science Institute, 776 Daedeokdae-ro, Yuseong-gu, Daejeon 34055, Korea}
.pdf

\author{Heeyoung Oh}
\affil{Korea Astronomy and Space Science Institute, 776 Daedeokdae-ro, Yuseong-gu, Daejeon 34055, Korea}

\author{Hwihyun Kim} 
\affil{Gemini Observatory, Casilla 603, La Serena, Chile}

\author{Benjamin T. Kidder}
\affiliation{Department of Astronomy, The University of Texas at Austin, Austin, TX, 78712, USA}

\author{Daniel T. Jaffe}
\affiliation{Department of Astronomy, The University of Texas at Austin, Austin, TX, 78712, USA}




\begin{abstract}

At 60 pc, TW Hydra (TW Hya) is the closest example of a star with a gas-rich protoplanetary disk,  though TW Hya may be relatively old (3-15 Myr). As such, TW Hya is especially appealing to test our understanding of the interplay between stellar and disk evolution. We present a high-resolution near-infrared spectrum of TW Hya obtained with the Immersion GRating INfrared Spectrometer (IGRINS) to re-evaluate the stellar parameters of TW Hya. We compare these data to synthetic spectra of magnetic stars produced by MoogStokes, and use sensitive spectral line profiles to probe the effective temperature, surface gravity, and magnetic field. A model with $T_{\rm eff} = 3800$ K, $\log \rm{g}=4.2$, and B$=3.0$ kG best fits the near-infrared spectrum of TW Hya. These results correspond to a spectral type of M0.5 and an age of 8 Myr, which is well past the median life of gaseous disks.

\end{abstract}





\section{Introduction}

Revolutionary new instruments have opened new windows into the detailed physical processes of star, disk, and planet formation and evolution. We have already witnessed one of the most anticipated achievements at millimeter wavelengths: to spatially resolve structure in protoplanetary disks. The first, HL Tau, was beautifully imaged by ALMA in early science with a spatial resolution of 0$\farcs$025, corresponding to 3.5 AU \citep{brogan15}. The protoplanetary disk of the young star HL Tau surprisingly exhibits clear rings throughout. \citet{awz15} obtained an even higher resolution (0$\farcs$02 $\sim$ 1AU) image of TW Hydra (TW Hya), host to one of the closest protoplanetary disks.  The star TW Hya, member of its namesake TW Hydra association, is the closest known gas-rich classical T-Tauri star \citep[distance of 59.5pc;][]{gaia_a,gaia_b}. Like HL Tau's disk, the disk around TW Hya has a rich structure of rings and a dark annulus at 1 AU that may be indicative of a planet  \citep{awz15,tsu16,vb17}.

A full understanding of the physical implications of the complex, yet similar morphology seen in the ALMA images will require better insight into the disks through more detailed  observations, especially observations of disk kinematics.  We must also have a significantly better understanding of the host stars, in particular reliable ages, as the stellar age sets the chronology for the disk. There is perhaps no better example of our lack of understanding about the longevity of a disk than the case of TW Hya. Studies of the TW Hydra association members suggest ages between 7-10 Myr  \citep[e.g. ][]{webb99,wein13, duc14, don16}. A similar age of $\sim$ 10 Myr is found for TW Hya itself from an optically derived spectral type of K7, determined first by \citet{herbig78}, in addition to its photometry and placement on HR diagram isochrones \citep{webb99,yjv05}.  Yet, the median lifetime of gaseous disks is much younger at $\sim$ 2 Myr \citep{ev09,wc11}.  Also, HL Tau resides in a cluster aged $\sim$ 1 Myr \citep{bri02}, and therefore it is difficult to reconcile the similarity of the two disks if TW Hya is one of the oldest remaining rich disks.  Despite its age, TW Hya still has the potential to form planets \citep{ber13}, making it one of the oldest protoplanetary disks.

There is a problem with posing the age-evolutionary state puzzle with TW Hya as one anchor, as the age of TW Hya is hotly debated. The catalyst was the first spectroscopic study of TW Hya at near-infrared wavelengths by \citet{vs11} (VS11). VS11 used medium resolution near-IR spectra and found a significantly later spectral type of M2.5, based on template fitting, than was typically concluded \citep[K7; e.g. ][]{herbig78}.  With this revised spectral type,  VS11 infer that TW Hya is younger (3 Myr). Various work since then has found spectral types and ages in between K7 and M2.5 and 3 and 10 Myr, or even simply have adopted an intermediate value, as in \citet{wein13}. For example, a study by \citet{debes13} determines that the optical to near-infrared data of TW Hya are best fit with a composite K7$+$M2 template. Alternatively, the recent work of \citet{hh14} argues that fitting a two temperature scheme is not necessary, finding instead that a template spectral type of M0.5 star corresponding to 3810K is representative of TW Hya (yet, these authors still find an old age of $\sim$ 15 Myr).

The uncertainty regarding TW Hya underlines our need for accurate stellar characterization for YSOs. However, observations of YSOs can be strongly influenced by features that come with their youth, such as surrounding dark clouds, accretion and strong magnetic fields. These attributes can manifest as reddening, veiling, and multiple surface temperature zones due to stellar spots. Veiling, an excess continuum emission that washes out the spectrum of the star by making the stellar absorption lines appear weaker by a factor depending upon wavelength, is an especially difficult observational challenge because it changes line equivalent widths. Most spectral analysis is, therefore, limited to using pairs of nearby spectral lines when relying on equivalent widths when working with moderate resolution spectra and line profiles when working with high resolution spectra (which can still be influenced by these factors) to try to avoid these observational challenges.

Observations at longer wavelengths can provide advantages over optical wavelengths. Particularly for cool stars like TW Hya, the color of (V $-$ K) is large, and
combined with the significant reddening in star-forming regions and the frequent presence of local obscuration around YSOs together make it more favorable to observe these sources in the infrared. Moreover, the temperature contrast between the stellar photosphere and the stellar spot is more severe in optical spectra.  In the infrared, the contrast between the star and stellar spot is less pronounced and is rather closer to an area-weighted mean spectrum. As such, infrared spectra are more representative of the true nature of the stellar surface by giving the average surface temperature. 
Lastly, infrared spectra present an opportunity to measure the magnetic field; the magnitude of Zeeman broadening increases with increasing wavelength as $\lambda^2$ whereas Doppler broadening only increases as $\lambda$. Thus, the shapes of infrared spectral lines can be more sensitive to the magnetic field strengths than lines in at optical wavelengths.

In this paper, we take advantage of near-infrared wavelengths to resolve the debate over the spectral type of TW Hya. We re-evaluate the stellar parameters of TW Hya using a powerful combination of high resolution near-infrared spectra and a spectral synthesis code that includes magnetic fields. We use MoogStokes to compute the emergent spectrum of a magnetic star \citep{deen13}. We present a new high signal-to-noise spectrum of TW Hya, obtained with the Immersion GRating INfrared Spectrometer \citep[IGRINS; ][]{park14,mace16}, which boasts both high resolving power ($R=\frac{\lambda}{\delta\lambda} = 45000$) and large spectral grasp (1.5-2.5 $\mu$m).  Armed with these models and the new data, we are able to determine more accurate values for effective temperature, surface gravity, and disk-averaged magnetic field strength for TW Hya.

\section{Observations and Data Reduction}

We observed TW Hya with IGRINS on the 2.7m Harlan J. Smith Telescope at McDonald Observatory in 2015 and 2017 (see Table \ref{table-obs} for details; airmass $=$ 2.4 for all observations) by nodding TW Hya along the slit in ABBA patterns.  Observations of standard A0V telluric stars were obtained in the same fashion for use in telluric corrections. 

\begin{figure*}
\includegraphics*[width=0.95\textwidth,angle=0]{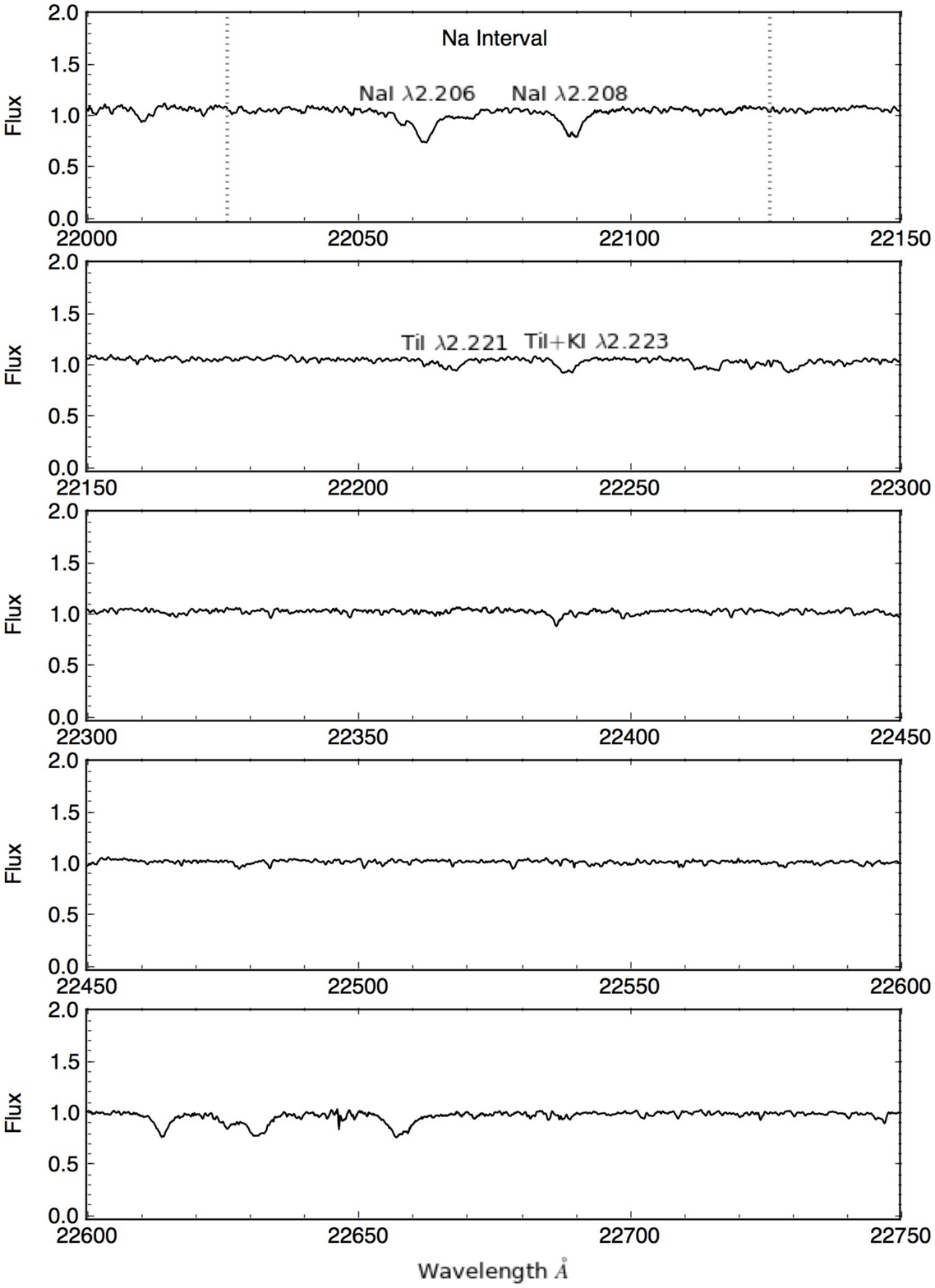}
\caption{\label{fig-kband-observed} The complete IGRINS spectrum of TW Hya from 1.45 $\mu$m to  2.5 $\mu$m is available in the online journal. In this example, the Na interval is marked by vertical dotted lines and labels, and the Ti lines used in the analysis are indicated.}
\end{figure*}

We use the IGRINS pipeline package \citep[version 2.1 alpha 3; ][]{lee15} to reduce the spectroscopic data, resulting in a one-dimensional, telluric-corrected spectrum with wavelength solutions derived from OH night sky emission lines at shorter wavelengths and telluric absorption lines longward of 2.2 $\mu$m. Telluric correction is performed by dividing the target spectrum by the A0V telluric star. We then combine the reduced spectroscopic data from each night by first correcting the wavelengths of each epoch for the barycenter velocity and calculate the flux as the median normalized flux from all epochs in 1-pixel (0.00001 $\mu$m) bins with uncertainties given by the standard deviation of the mean. The flux is normalized by dividing by the median flux over a wavelength window (from 1.715 -- 1.75 $\mu$m for the H-band and 2.25 -- 2.29 $\mu$m for the K-band spectra). The final combined spectrum has a median signal-to-noise ratio of 50 in the H-band and 60 in the K-band.  The entire IGRINS spectrum from 1.45 $\mu$m to 2.5 $\mu$m is shown in Figure \ref{fig-kband-observed} in the online journal; although the analysis in this paper focuses on spectral regions in the K-band. A strong Br$\gamma$ emission line at 2.17$\mu$m is quite noticeable, shown in Figure 1.9; weaker Bracket emission features (Br10 and Br11 in Figure 1.3) are also clearly present.

\figsetstart
\figsetnum{1}
\figsettitle{IGRINS spectrum of TW Hya}

\figsetgrpstart
\figsetgrpnum{1.1}
\figsetgrptitle{The first section of the IGRINS spectrum of TW Hya, starting at $\lambda$ 1.45 $\mu$m.}
\figsetplot{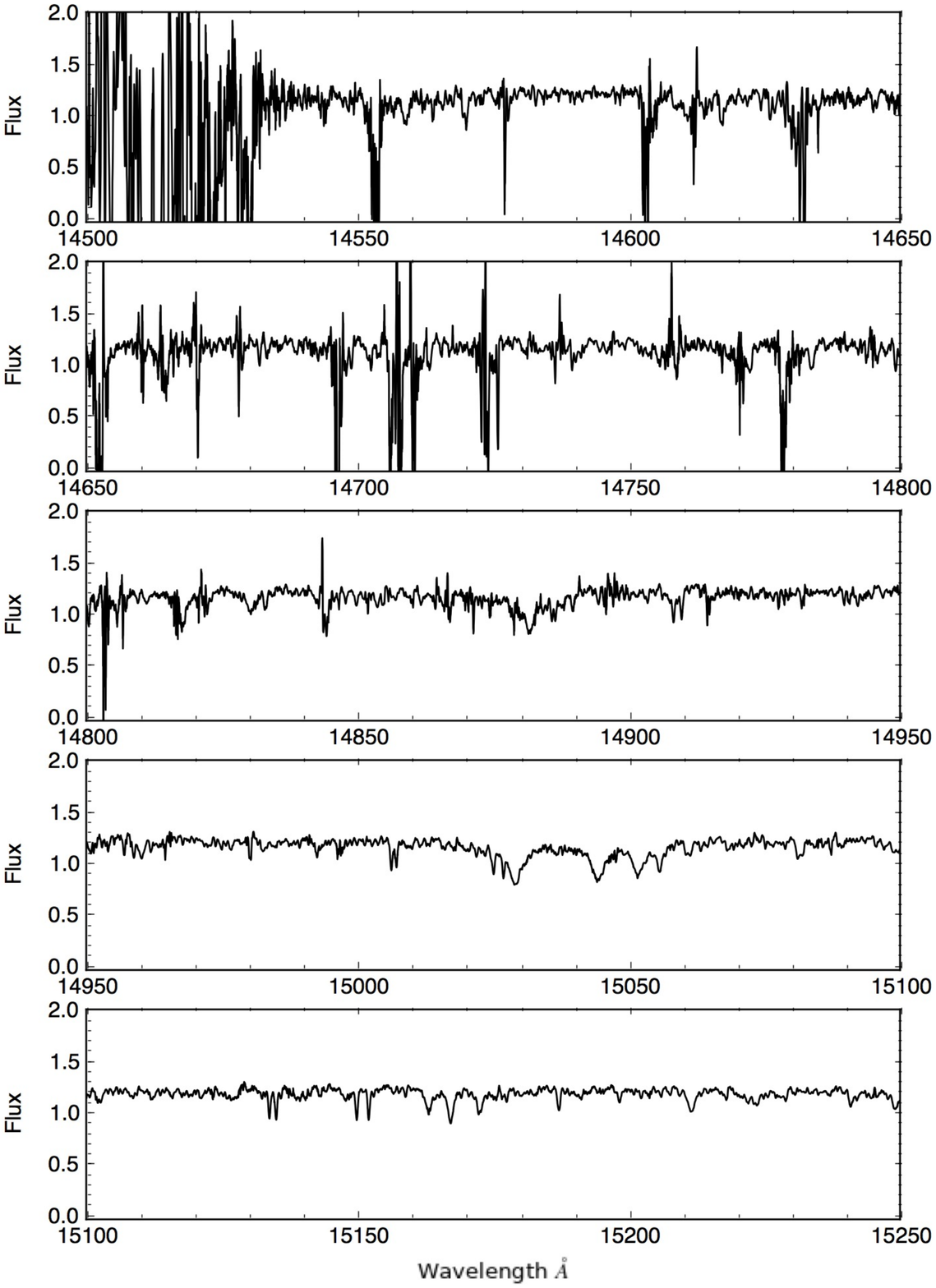}
\figsetgrpnote{ }
\figsetgrpend

\figsetgrpstart
\figsetgrpnum{1.2}
\figsetgrptitle{A continuation of the IGRINS spectrum of TW Hya.}
\figsetplot{PUB_spectra_full_H1.pdf}
\figsetgrpnote{}
\figsetgrpend

\figsetgrpstart
\figsetgrpnum{1.3}
\figsetgrptitle{A continuation of the IGRINS spectrum of TW Hya.}
\figsetplot{PUB_spectra_full_H2.pdf}
\figsetgrpnote{In this section, Bracket emission features are clearly present and labelled.}
\figsetgrpend

\figsetgrpstart
\figsetgrpnum{1.4}
\figsetgrptitle{A continuation of the IGRINS spectrum of TW Hya.}
\figsetplot{PUB_spectra_full_H3.pdf}
\figsetgrpnote{}
\figsetgrpend

\figsetgrpstart
\figsetgrpnum{1.5}
\figsetgrptitle{A continuation of the IGRINS spectrum of TW Hya.}
\figsetplot{PUB_spectra_full_H4.pdf}
\figsetgrpnote{}
\figsetgrpend

\figsetgrpstart
\figsetgrpnum{1.6}
\figsetgrptitle{A continuation of the IGRINS spectrum of TW Hya.}
\figsetplot{PUB_spectra_full0.pdf}
\figsetgrpnote{}
\figsetgrpend

\figsetgrpstart
\figsetgrpnum{1.7}
\figsetgrptitle{A continuation of the IGRINS spectrum of TW Hya.}
\figsetplot{PUB_spectra_full1.pdf}
\figsetgrpnote{}
\figsetgrpend

\figsetgrpstart
\figsetgrpnum{1.8}
\figsetgrptitle{A continuation of the IGRINS spectrum of TW Hya.}
\figsetplot{PUB_spectra_full2.pdf}
\figsetgrpnote{}
\figsetgrpend

\figsetgrpstart
\figsetgrpnum{1.9}
\figsetgrptitle{A continuation of the IGRINS spectrum of TW Hya.}
\figsetplot{PUB_spectra_full3.pdf}
\figsetgrpnote{In this section, the strong Br$\gamma$ emission line is labelled. }
\figsetgrpend

\figsetgrpstart
\figsetgrpnum{1.10}
\figsetgrptitle{A continuation of the IGRINS spectrum of TW Hya.}
\figsetplot{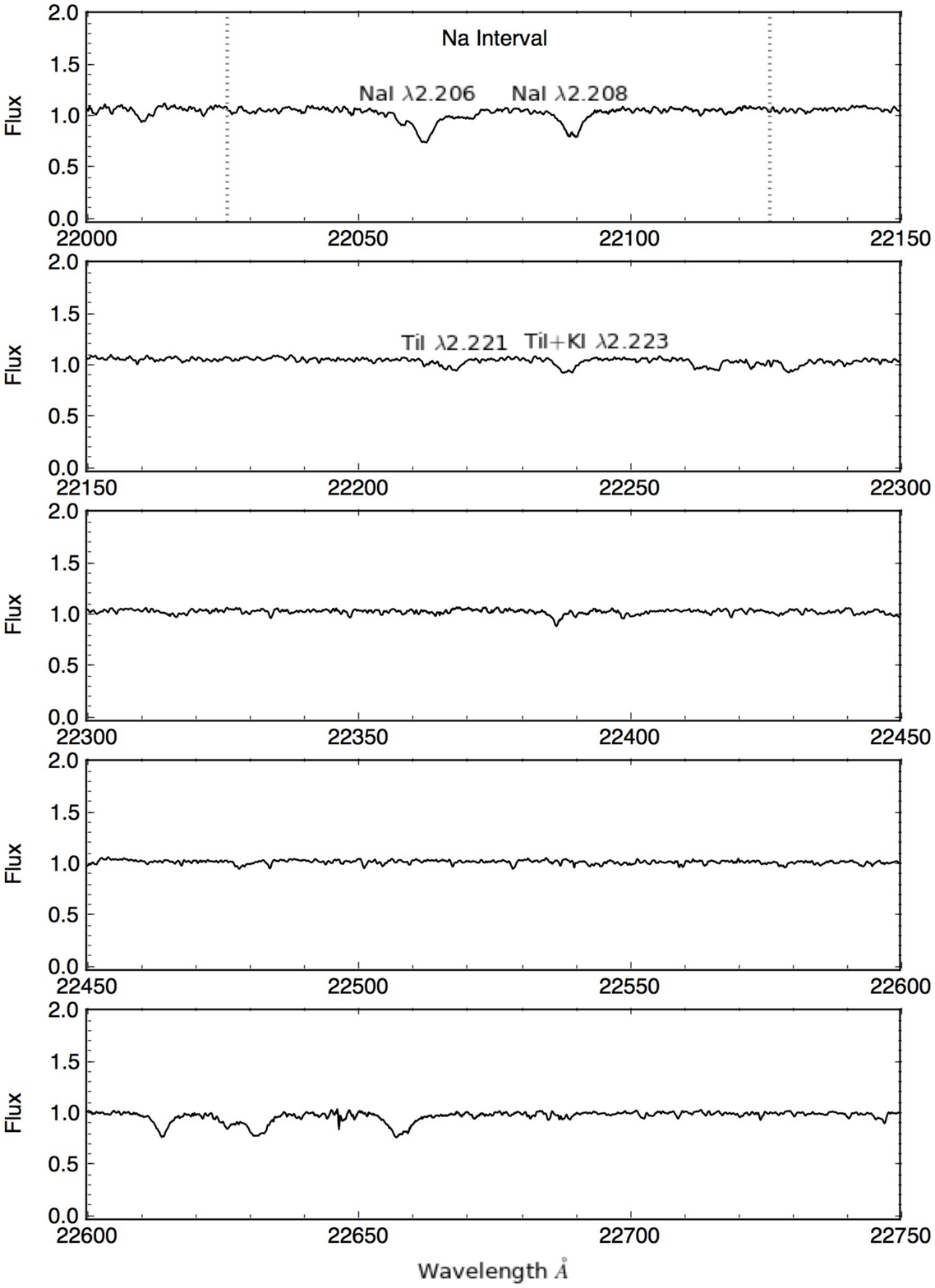}
\figsetgrpnote{In this section, the Na interval is marked by vertical dotted lines and labels, and the Ti lines used in the analysis are indicated.}
\figsetgrpend

\figsetgrpstart
\figsetgrpnum{1.11}
\figsetgrptitle{A continuation of the IGRINS spectrum of TW Hya.}
\figsetplot{PUB_spectra_full5.pdf}
\figsetgrpnote{In this section,  the (2-0) $^{12}$CO interval is marked by vertical dotted lines and labels.}
\figsetgrpend

\figsetgrpstart
\figsetgrpnum{1.12}
\figsetgrptitle{A continuation of the IGRINS spectrum of TW Hya.}
\figsetplot{PUB_spectra_full6.pdf}
\figsetgrpnote{}
\figsetgrpend

\figsetgrpstart
\figsetgrpnum{1.13}
\figsetgrptitle{The final section of the IGRINS spectrum of TW Hya.}
\figsetplot{PUB_spectra_full7.pdf}
\figsetgrpnote{}
\figsetgrpend

\figsetend

\begin{deluxetable}{lll}
\tabletypesize{\footnotesize}
\tablecolumns{8} 
\tablewidth{0pt}
 \tablecaption{IGRINS Observations of TW Hya \label{table-obs}}
 \tablehead{
\colhead{UT Date} & \colhead{Integration Time} & \colhead{A0V Calibrator} } 

 \startdata 
2015-01-06 & 240s $\times$ ABBA & HD 92845 \\ 
2015-01-28 & 240s $\times$ ABBAABBA & HD 89213 \\
2017-03-12 & 300s $\times$ ABBA & HR 3646 \\
2017-03-13 & 300s $\times$ ABBA & HR 5167 \\
\enddata

 \vspace{-0.8cm}
\end{deluxetable}

\section{Grid Synthetic Spectra from MoogStokes}

We generate a 3-dimensional grid (across the parameters of effective temperature $T_{\rm eff}$, surface gravity $\log {\rm g} $, and magnetic field strength B) of synthetic spectra using the MoogStokes code \citep{deen13}. MoogStokes, a customization of the one-dimensional LTE radiative transfer code Moog \citep{sne73}, synthesizes the emergent spectra of stars with magnetic fields in their photospheres. It assumes a uniform effective temperature and surface gravity and a uniform, purely radial magnetic field. While these assumptions are clearly non-physical and are not valid for accurate disk-resolved spectra of all the Stokes Q, U, and V components, they produce sufficiently accurate disk-averaged Stokes I spectra, which is  the spectra that can be measured by IGRINS. MoogStokes calculates the Zeeman splitting of an absorption line by using the spectroscopic terms of the upper and lower state to determine the number, wavelength shift, and polarization of components into which it will split for a given magnetic field strength. Two axes of the grid (effective temperature $T_{\rm eff}$, surface gravity $\log {\rm g} $) are defined by the model atmospheres; we use the solar metallicity \citep[appropriate for YSOs; ][]{padgett96,santos08} MARCS model atmospheres \citep{gus08} for this investigation. The third axis of the grid (mean magnetic field B) is input as desired.  

For each grid point, MoogStokes produces a suite of raw output composed of emergent spectra synthesized at seven different viewing angles across the stellar disk. The program uses the resultant raw output to generate a disk averaged synthetic spectrum after it applies the effects of limb darkening and rotational broadening given the source geometry \citep{deen13}. To compare to our target, TW Hya, we fix $v \sin i$ to the value in the literature, $v \sin i$  $=$ 5.8 km/s \citep{yjv05, ab02}. To then enable a direct comparison to our IGRINS spectra, we convolve the synthetic spectra with a gaussian kernel to simulate the $R=45000$ resolving power of IGRINS; additionally, we sample the synthetic spectra to simulate the digitization of the spectra by pixels of finite size. When we need a theoretical spectrum with $T_{\rm eff}$ or $\log {\rm g} $ values that lie between the values of the published MARCS models, we linearly interpolate between synthetic spectra for parameter values in between grid points.

\section{Spectral Analysis}

\subsection{Highlighting Sensitive Spectral Features}

\begin{figure}[t!]
\includegraphics*[width=0.45\textwidth,angle=0]{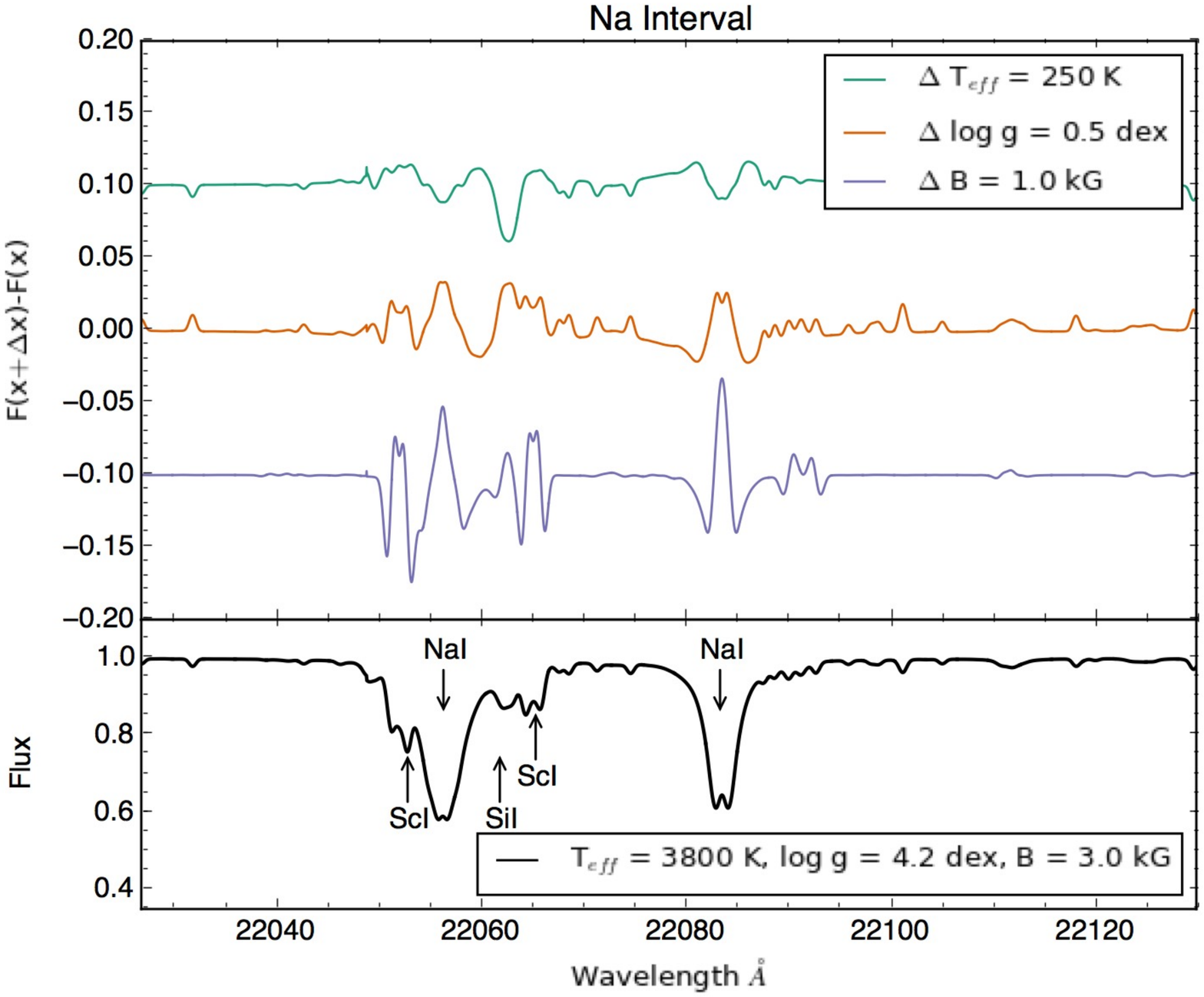}
\includegraphics*[width=0.45\textwidth,angle=0]{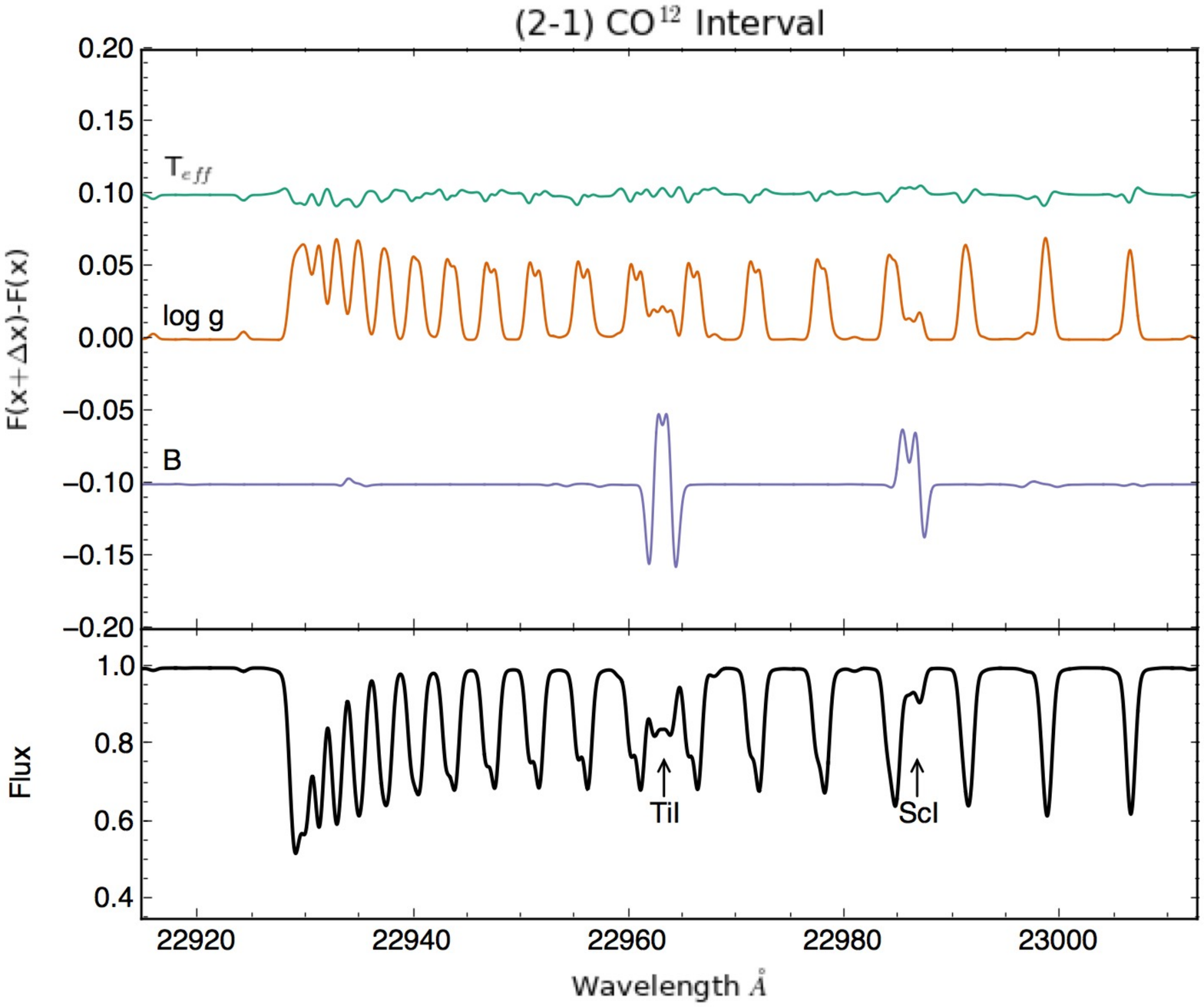}
\includegraphics*[width=0.45\textwidth,angle=0]{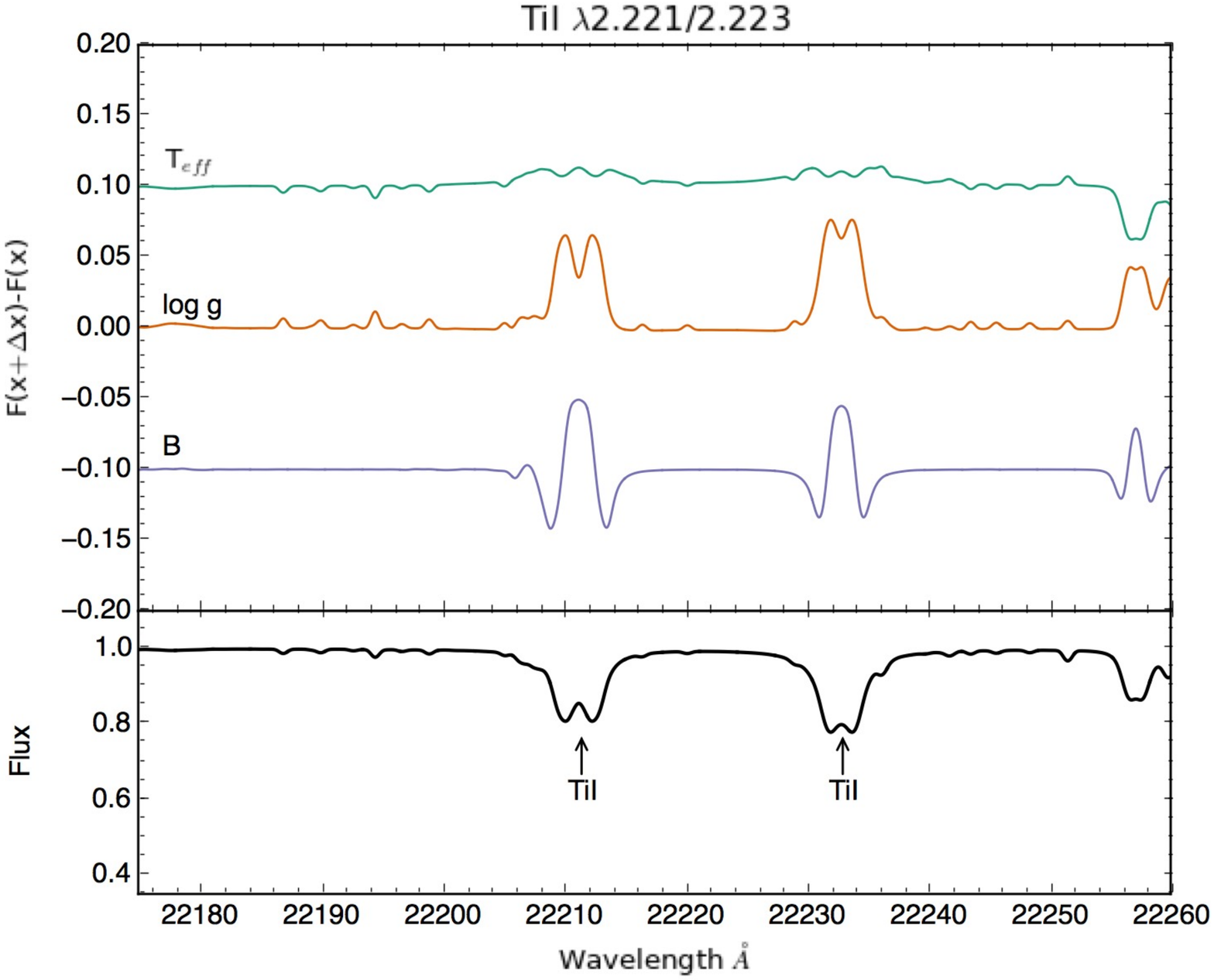}
\caption{\label{fig-intervals-diffsens} Parameter sensitivity of R= 45,000 Moog Stokes spectral synthesis models for three K-band spectral regions.  The bottom portion of all three panels shows the MoogStokes spectrum for a fiducial model with $T_{\rm eff} = 3800 $K, $\log \rm{g}=4.2$, and B $=3.0 $ kG.  The top portion of each panel shows the difference in the synthetic spectra caused by changing $T_{\rm eff} $ by 250 K, $\log \rm{g}$ by 0.5, and B by 1.0 kG.  In each case, we computed this difference by holding two of the parameters fixed and varying the third up and down from the fiducial value by the given amount ($\Delta$) and taking the average of the spectral change resulting by a decrement and an increment. }
\vspace{-30pt}
\end{figure}

The sensitivity of the shapes and strengths of the spectral lines present in the observations of TW Hya in the 2.10-2.45 $\mu$m spectral region to effective temperature, surface gravity, and magnetic field strength give us the ability to measure all three parameters. We evaluate and show examples of the change in line profiles through differentials of the MoogStokes models. The differentials represent the average change in flux from models with  a positive and negative step of size $\Delta$ for an individual parameter.  We center our differential illustration on a base model of $T_{\rm eff} = 3800 $K, $\log \rm{g}=4.2$, and B$=3.0 $ kG, and vary one of the parameters from those starting values while holding the other parameters constant.  Figure \ref{fig-intervals-diffsens}, which is composed of three subfigures of specific spectral regions, shows the differentials pertaining to each parameter plotted in the top panel (colored lines) and the fiducial flux of the baseline spectral model in the bottom panel. The illustration in Figure \ref{fig-intervals-diffsens} uses step sizes that correspond to the maximum step sizes defining the model grid. 

The differentials shown in Figure \ref{fig-intervals-diffsens} serve as a guide to the parameter values that may be important in shaping the absorption lines of TW Hya  and to develope intuition for the parameter(s) responsible for altering line profiles in this part of $T_{\rm eff}$, $\log \rm{g}$, B space. Using this guide, we identify these spectral regions to later determine the stellar parameters (Section \ref{sec-parameters}), which are the Na interval (2.202-2.212 $\mu$m), the (2-0) $^{12}$CO interval (2.2925-2.3022 $\mu$m), and Ti lines near 2.221$\mu$m and 2.223 $\mu$m (using an interval over 2.220-2.224 $\mu$m). Lines from other species are also present in these intervals. The Na and $^{12}$CO intervals are similar to those given by  \citet{dj03}, and these Ti lines have been used for examining the strength of the magnetic field, for instance by \citet{yjv05}

The sensitivity of the CO interval (middle of Fig. \ref{fig-intervals-diffsens}) to the surface gravity is the most obvious result of this exercise. The differential pertaining to a step in the magnetic field parameter is markedly flat  across this interval except for two distinct features (a Ti line and a Sc line); furthermore, the changes due to varying the effective temperature are insignificant in comparison to those induced by the surface gravity.  Many other key features can be easily seen throughout these differentials and are discussed further in Section \ref{sec-parameters}. 

\subsection{\label{sec-parameters} Parameterization of TW Hya}

\subsubsection{Identifying the Best Fit Synthetic Spectrum}

To characterize the physical parameters of TW Hya, we implemented a hands-on, iterative  process to find the best MoogStokes model to match the observed IGRINS spectrum. 
We identify the best fitting model by varying three stellar parameters: effective temperature, surface gravity, and magnetic field strength. We hold the values of two of the parameters constant, and then compare the synthetic spectra resulting from a range of values of the third parameter to the observed spectrum. The best value for that parameter is then found, adopted, and set; then the same process for the next parameter begins. This process is repeated over the three parameters until convergence is observed. 

The resulting method restricts analysis to optimal spectral regions that -- outlined in the section above -- are strongly sensitive to just one of the stellar parameters: effective temperature, surface gravity, or magnetic field strength.  We first used the approach of \citet{dj03} as a guide; they presented a clear, broad-brush iterative method to estimate the effective temperature and surface gravity of a YSO using the Na and $^{12}$CO intervals.  We add iterations over the magnetic field strength using the Ti lines  near 2.221$\mu$m and 2.223 $\mu$m as an additional parameter.  Our evaluation of the best fit model is done in a similar manner as to \citet{mohanty04}, who  also use narrow wavelength ranges at high spectral resolution (although they do not iterate for a final solution). \citet{mohanty04} identify the best fit model first visually, then verify through a goodness of fit minimization. We cycle through identifying the best fitting value for each of the parameters to converge on the final best fit model, and ultimately use goodness of fit minimization to define the uncertainties, discussed in the following section.
 
Before the fit between an observed IGRINS spectrum and the MoogStokes synthetic spectra can be evaluated, the observed spectrum must be flattened and the synthetic spectra must be artificially veiled such that all the spectra can be compared in the same format. The observed spectrum is flattened and normalized using an interactive python script (based on \url{http://python4esac.github.io/plotting/specnorm.html}). When defining the estimated continuum level, we looked at the variation from night to night as an estimator for noise. Then to artificially veil the synthetic spectra, the value of the veiling (r$_{k}$) must be found. For each synthetic model, we measure the best fit veiling value to the observed spectrum  at 2.2 $\mu$m using a least squares fitting routine. By measuring the veiling at 2.2 $\mu$m, we are using a different section of the spectrum than is used to determine the stellar parameters, minimizing the degeneracy. Then, each individual synthetic spectrum is artificially veiled by the measured value r$_{k}$ and assuming blackbody emission at $\sim$ 1500 K due to a warm dust component  \citep{cieza05} (rather than accretion that can cause veiling effects at shorter wavelengths). The result is a unique measurement of the veiling for every model that we compare to; the final adopted veiling measurement corresponds to that found with the best fitting model.

The best fit value of the parameter being tested is evaluated by comparing the synthetic spectra to the observed spectrum both by the shape of the line profile(s) by eye and the computed root-mean-square (rms) across a given spectral region. When varying a parameter, a broad range of values across that parameter space is first adopted, then smaller steps are taken,  e.g. the two-step approach as recently discussed by \citet{ec16}. The final step sizes used are $T_{\rm eff} = 100 K$, $\log {\rm g} = 0.1$, and B $ = 0.1$kG.

\begin{figure*}[h!]
\includegraphics*[width=0.95\textwidth,angle=0]{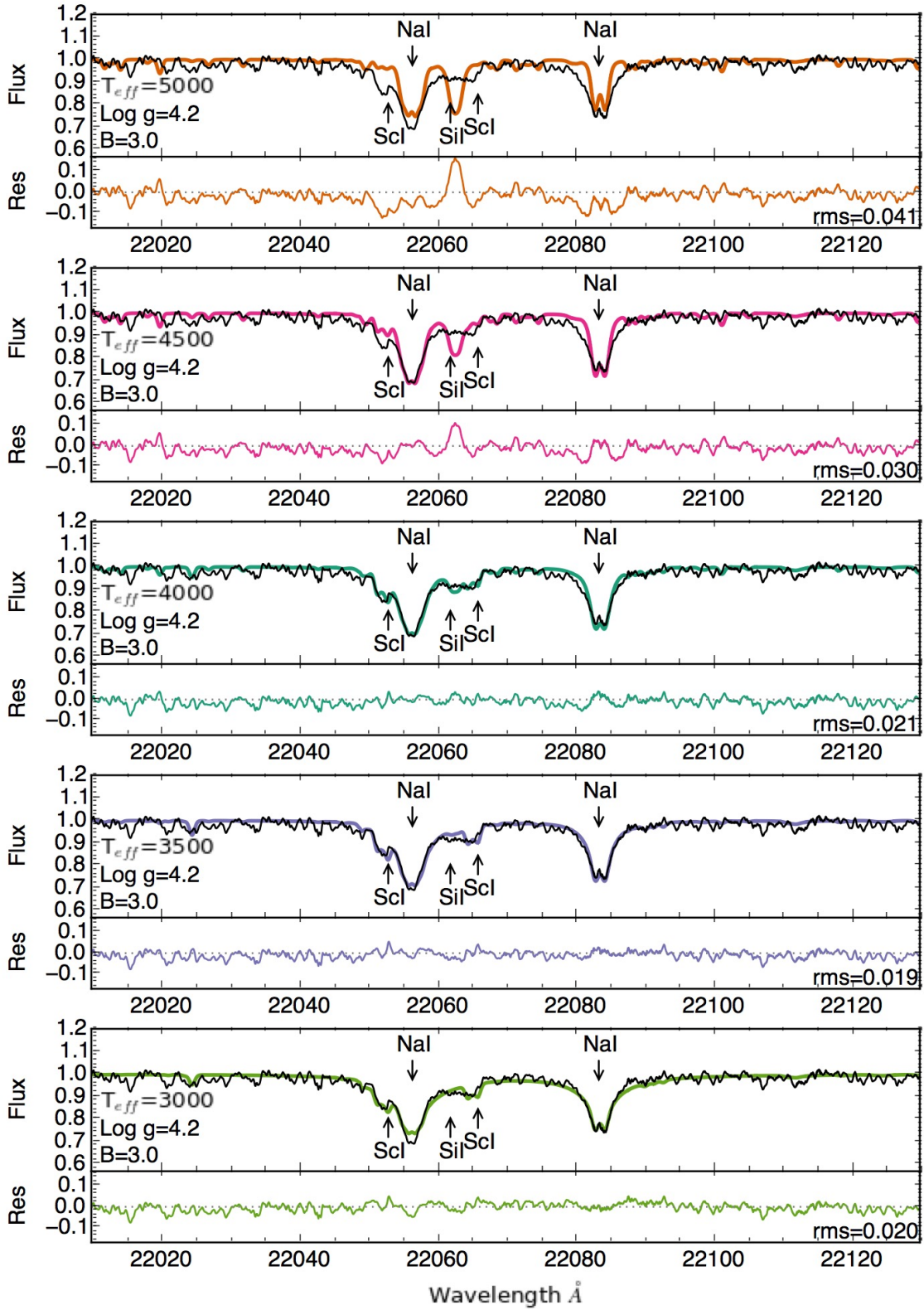}
\caption{\label{fig-rms-temp} A plot illustrating a step in the iterative process of identifying the stellar parameters of TW Hya. Each panel shows a different MoogStokes model (colored lines, descriptions given in each respective panel) in comparison to the observed IGRINS spectrum (black line). The residuals are shown in the bottom portion of each panel (with an expanded y-axis scale). 
For this illustrative example, the effective temperature is varied over the full range of the model grid, while the surface gravity and magnetic field are fixed to the best-fit value, and the resulting models are evaluated over the Na interval.  }
\end{figure*}

We begin by estimating effective temperature, using a fixed surface gravity and magnetic field strength, then determine the value of $T_{\rm eff} $ from  the synthetic spectrum that best matches the observed spectrum. Comparing the observed IGRINS spectrum of TW Hya to the MoogStokes models at a fixed surface gravity and magnetic field strength, it is clear that the strongest effective temperature indicator is the Si/Sc line ratio in the Na interval (see Figure \ref{fig-rms-temp}), with some sensitivity in the wings of the 2.208 $\mu$m Na line and the core of the 2.206 $\mu$m Na line. While a reasonable match to the observed Sc/Si line ratio can be found with MoogStokes models at the lowest  effective temperatures (Figure \ref{fig-rms-temp}), the poor fit to these low effective temperatures is exhibited by how depth of the observed Na lines cannot be matched at the same time. Agreement for the Na core depth and the Sc/Si line ratio is only achieved with models corresponding to higher effective temperatures.  

\begin{figure*}[t!]
\includegraphics*[width=0.95\textwidth,angle=0]{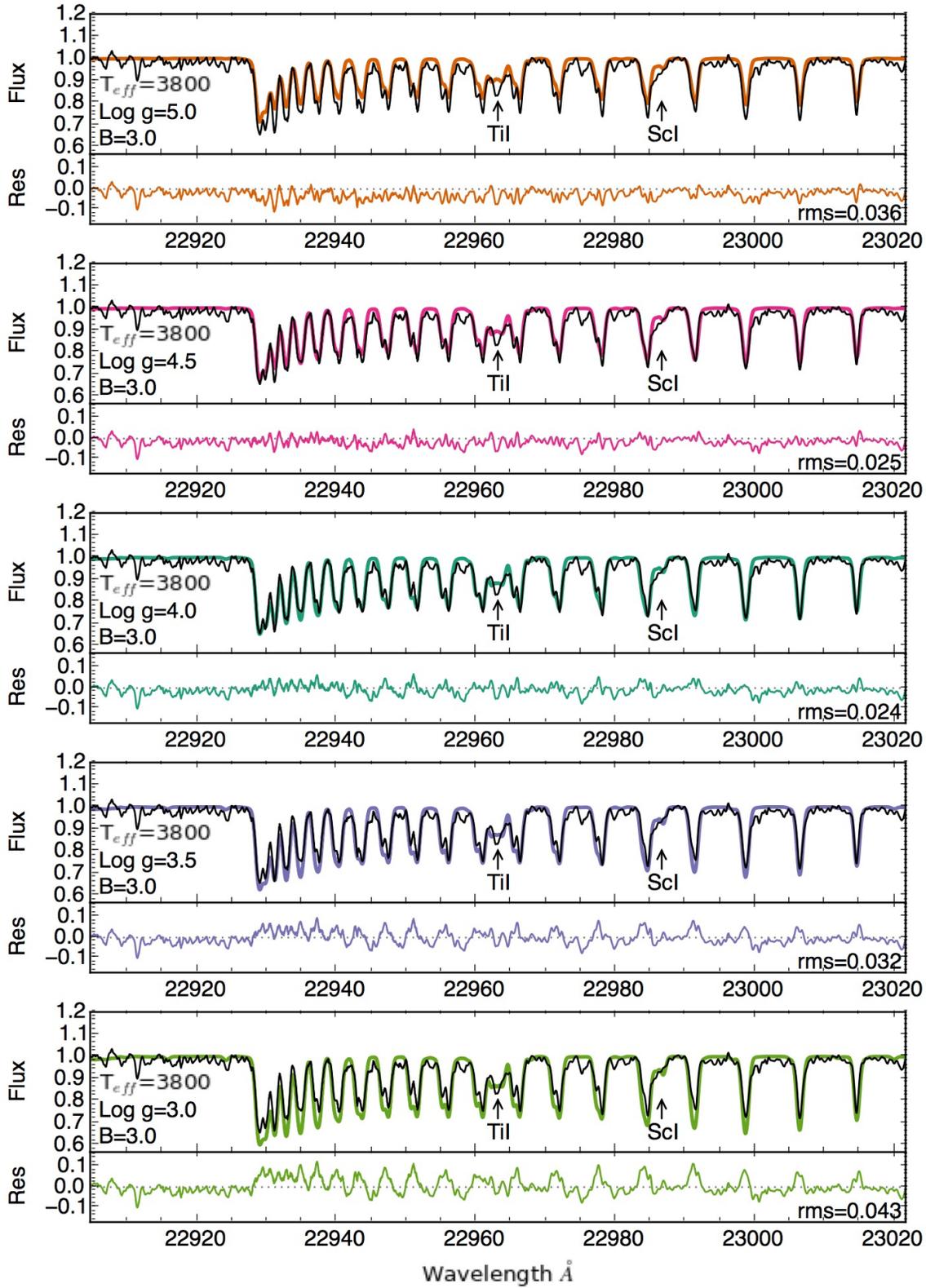}
\caption{\label{fig-rms-grav} Similar to Figure \ref{fig-rms-temp}. For this example, the surface gravity is varied while the effective temperature and magnetic field are held constant at the best fit value. The CO interval is shown.}
\vspace{35pt}
\end{figure*}

\begin{figure*}[t!]
\includegraphics*[width=0.95\textwidth,angle=0]{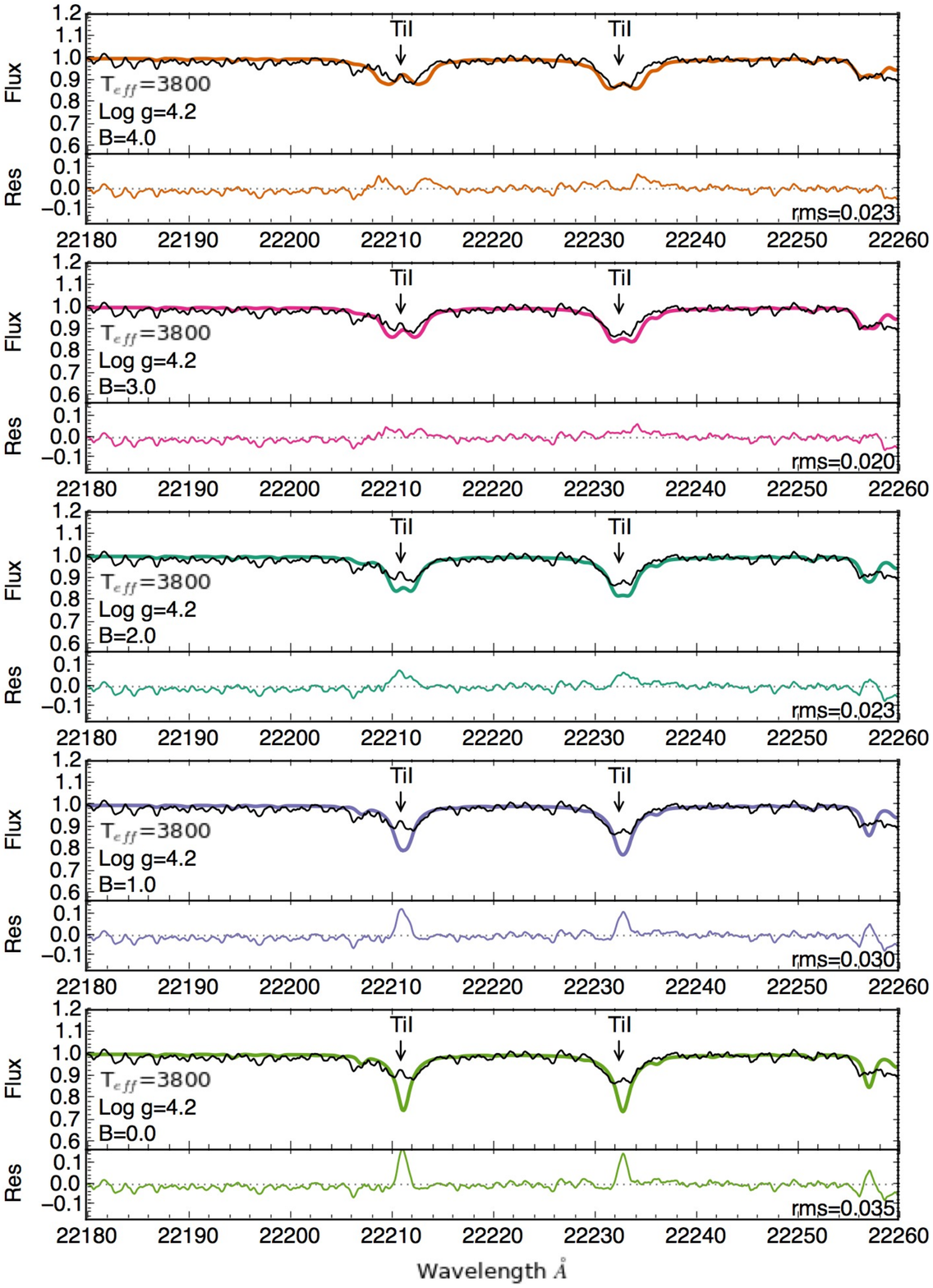}
\caption{\label{fig-rms-mag} Similar to Figure \ref{fig-rms-temp}. For this example, the magnetic field is varied over the full range of the model grid, while the effective temperature and surface gravity are fixed to the best-fit value, and the resulting models are evaluated over the magnetically sensitive Ti lines. }
\vspace{10pt}
\end{figure*}

After deriving an initial $T_{\rm eff}$, we examine the value of the surface gravity by fixing the effective temperature to its new optimal value and leaving the initial guess for the magnetic field unchanged. In determining the surface gravity, a comparison between the observed spectrum and the models was made between spectral features  present in the (2-0) $^{12}$CO interval. Figure \ref{fig-intervals-diffsens}  shows that the first bandhead, contained within the $^{12}$CO interval, and the three lines at the longest wavelengths within the interval (which are individual low J components) are strongly affected by the surface gravity. These lines are less sensitive to changes in the effective temperature, and therefore more restrictive tests of the surface gravity.  Additionally, while CO line strength is degenerate between veiling and surface gravity, also discussed in Section \ref{sec-comparison}, our independent measurement of the veiling (found at 2.2$\mu$m and assuming blackbody emission) breaks this degeneracy. By first finding the value of the veiling at a different section of the spectrum and adopting blackbody emission, the veiling is therefore already accounted for before diagnosing the $^{12}$CO interval.

After identifying the best fitting value of surface gravity, it is adopted, and the value of the magnetic field is then varied and evaluated. The strength of the magnetic field can have a fairly obvious effect. The bottom subfigure of Figure \ref{fig-intervals-diffsens} shows that shape of the core of the Ti lines  near 2.221$\mu$m and 2.223 $\mu$m can be strongly affected by the magnetic field (Zeeman splitting), and that this effect can dominate over changes of the other two parameters. Other examples of Zeeman splitting in these specific lines has been observed in other stars as well, even a Class I protostar \citep{jk09}. In addition to the line splits in the observed Ti lines in the IGRINS spectrum of TW Hya, the cores of the Na lines in the Na interval show some splitting. These features are clear evidence of a strong magnetic field in TW Hya. 
  
Examples of the described method are shown in Figures \ref{fig-rms-temp}-\ref{fig-rms-mag}, which display the models produced by varying one of the following parameters while holding the other two constant: the effective temperature, surface gravity, and magnetic field strength. For illustrative purposes, we show the variations of a given parameter when the fixed parameters are set to the final best fit values; these values are discussed further in Section \ref{sec-results}.

\subsubsection{Uncertainties via the Monte Carlo Method}

 We estimate our uncertainties by performing a Monte Carlo simulation to account for the impact of our observed flux uncertainties on finding the best fit synthetic spectra. We employ the goodness of fit statistic \citep[as in ][]{cush08} measured in specific spectral regions in order to determine best fitting synthetic spectra,  as is rather common \citep[e.g. ][]{stelzer13, man17}. Based on the knowledge gained from the method described above, we constructed a goodness of fit minimization algorithm and then identify our uncertainties with the Monte Carlo technique, following \citet{cush08, rice10, malo14}. The Monte Carlo is performed by constructing a simulated spectrum from flux values randomly sampled at each wavelength from a Gaussian distribution centered on the observed value with the width of the observed uncertainty, and then finding the best fitting synthetic spectrum via the minimum goodness of fit statistic. 

This program mimics our iterative procedure above to find the best fit for each simulated spectrum: varying the effective temperature using small steps across the grid while holding the surface gravity and magnetic field strength constant, then adopting the value of the effective temperature from the model with the best goodness of fit value, and then repeating this process with the adopted value to find the surface gravity and then the magnetic field values. This cycle is repeated, iterating over the effective temperature, surface gravity, and magnetic field parameters until the model with the best goodness of fit is identified. As before, specific spectral regions  are used to characterize specific parameters: across the Sc and Si line in the Na interval for the effective temperature, the $^{12}$CO interval for the surface gravity, and the Ti interval for the strength of the magnetic field. The goodness of fit measured over the Na interval would get stuck at the lowest effective temperature values, which clearly do not provide the best fit to the observed data (see Figure \ref{fig-rms-temp}), and therefore a narrower region focusing on the more sensitive Sc and Si lines (2.2060-2.2067 $\mu$m) was used. The initial guesses were kept constant and were the parameter values corresponding to the final best fit model (Section \ref{sec-results}). The veiling was also set to the value found for the final best fitting model (r$_{k} =$ 0.4).

We generated 250 simulated, randomly-sampled spectra. The distribution of the best fitting parameters suggest uncertainties well below the model grid spacing  ($\sim$ a factor of 10 less) and the peak values agreed with the final best fit value for the observed spectrum. Therefore, we adopt the step sizes of each parameter used in determining the best fit as the uncertainty (100K and 0.1 for effective temperature and surface gravity, respectively), except for the magnetic field. While the mean of the distribution of the best fitting magnetic field agrees with the final best fit value of the observed spectrum, the shape was bimodal with peaks at $\pm$ 0.1kG from the center, and we instead adopt an uncertainty of 0.2 kG for the magnetic field strength.

\subsubsection{Results \label{sec-results}}

\begin{figure*}[t!]
\includegraphics*[width=0.33\textwidth,angle=0]{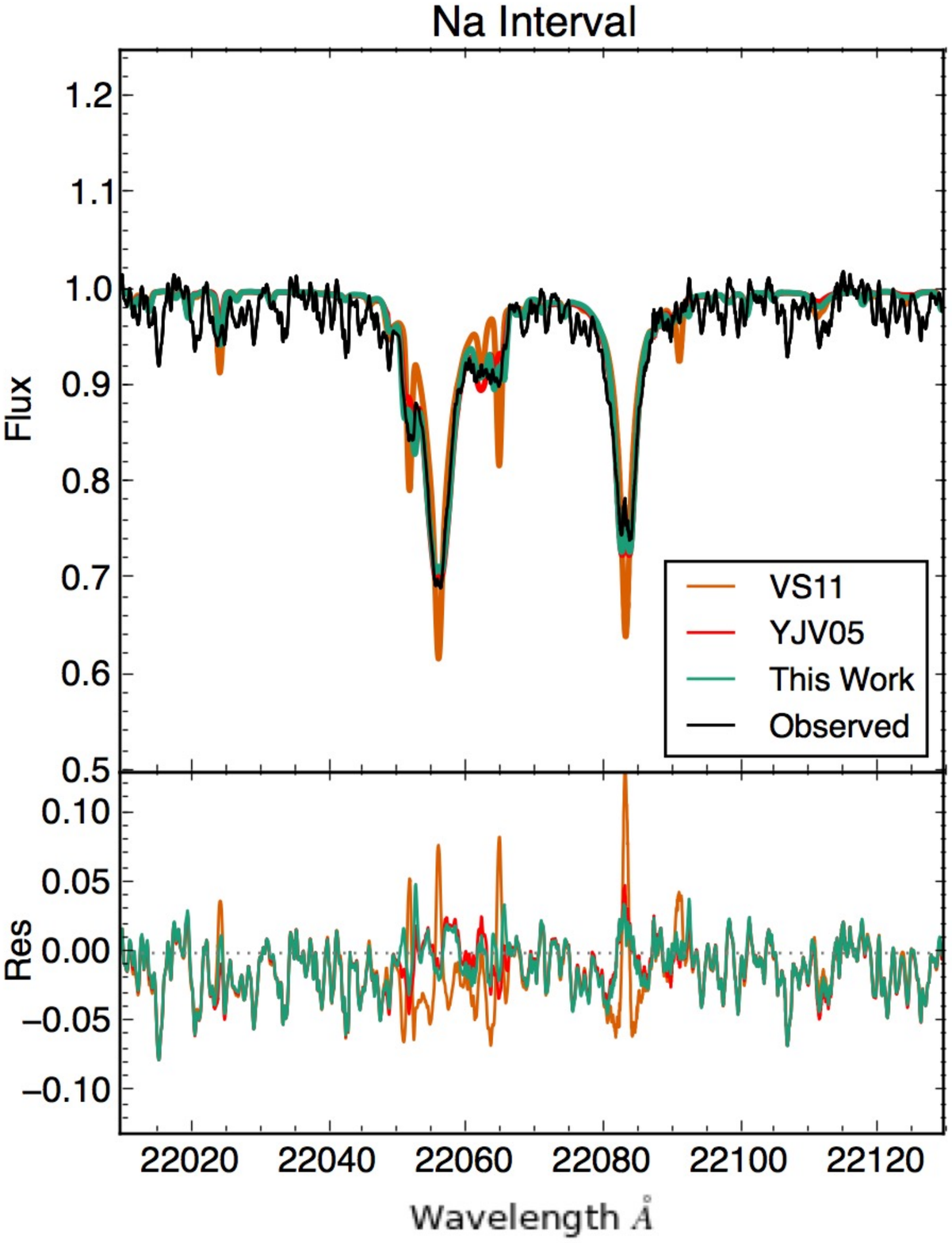}
\includegraphics*[width=0.33\textwidth,angle=0]{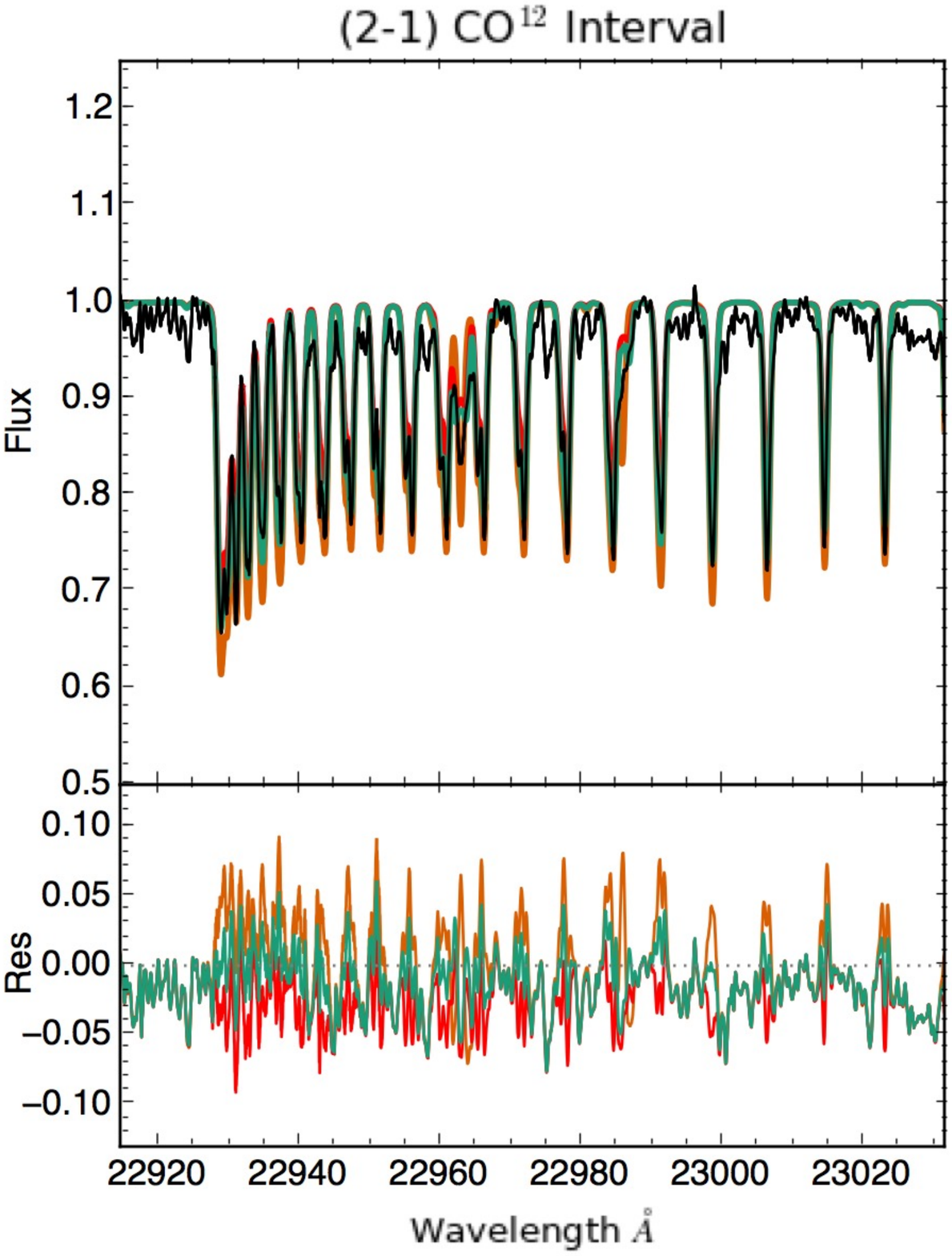}
\includegraphics*[width=0.33\textwidth,angle=0]{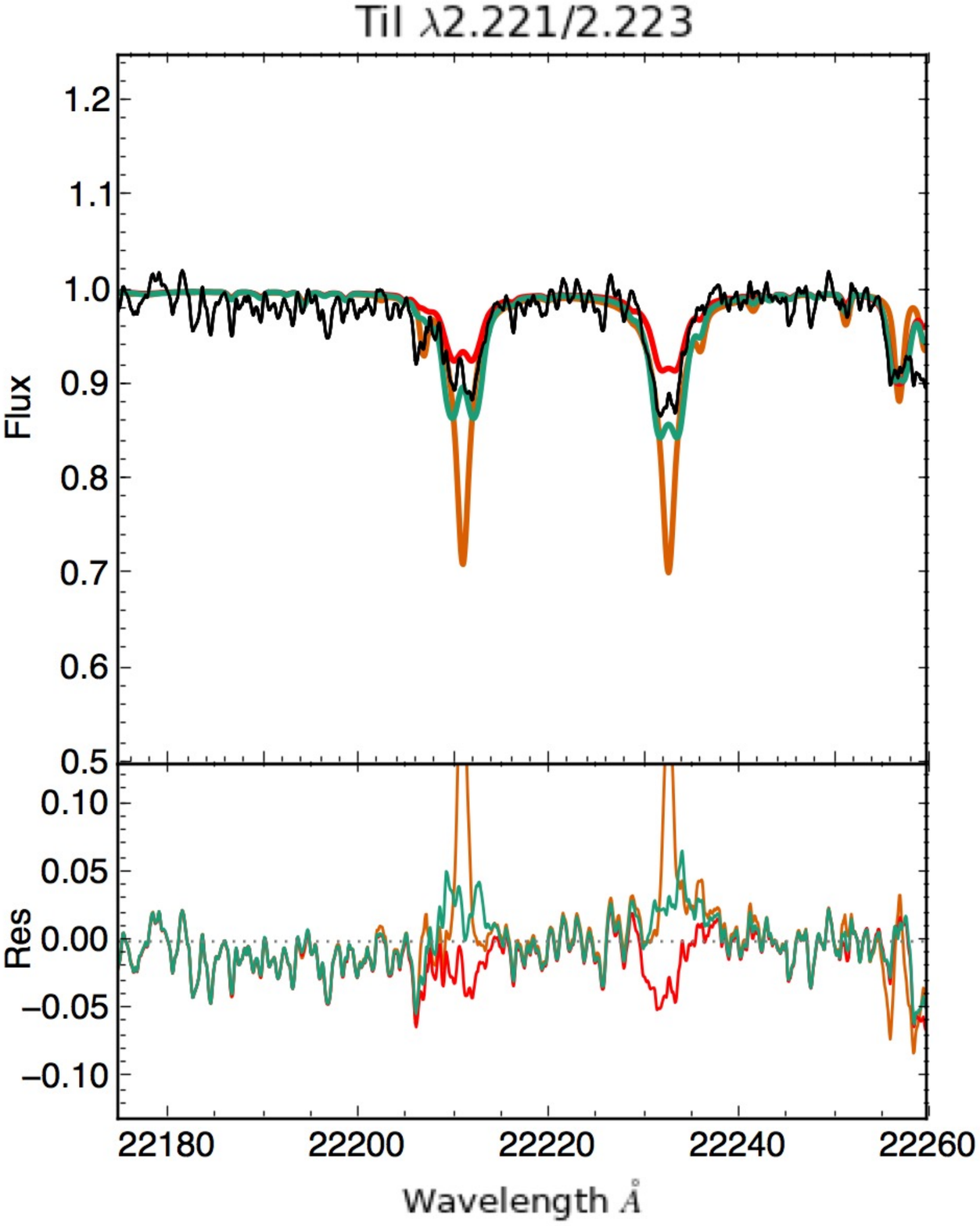}

\caption{\label{fig-blended} The observed IGRINS spectrum (black line) of TW Hya in comparison to MoogStokes synthetic spectra matching the estimated parameters of this paper,YJV05, and VS11 (colored lines; parameters are given in Table \ref{table-compare}). The difference between the observed and synthetic spectra is shown in the bottom panels, with an increased y-axis scale. Left: the Na interval, center: the $^{12}$CO interval, right:  Ti lines. A full view of the K-band spectrum is included in the online version (Figure \ref{fig-kband-observed}).}
\end{figure*}

We find the best fit for the IGRINS observations of TW Hya is produced by a MoogStokes synthetic spectral model with the stellar parameters of $T_{\rm eff} = 3800 \pm100$ K, $\log \rm{g}=4.2 \pm 0.1$, and B$=3.0 \pm 0.2$ kG. The veiling is r$_{k} \sim$ 0.4 at 2.2$\mu$m.   The best fit model is shown in Figure \ref{fig-blended}, and a comparison of models representing other reported results (discussed below) on top of the IGRINS data. The residuals between the models and the observed data are shown in the bottom panel.

\section{Discussion and Conclusions}

\subsection{Comparison to Other Studies of TW Hya \label{sec-comparison}}

The stellar parameters determined in this work from the IGRINS K-band spectrum fall firmly within previous results. Using the spectral type to temperature conversions of \citet{hh14}, we conclude that TW Hya is an M0.5, which is in between the typical optical spectral typing (K7) and the near-IR spectral typing of VS11 (M2.5). The literature shows variations of about $\sim$ 150K depending on the adopted spectral type to effective temperature conversion \citep{hh14}. The best agreement in the literature with our determined effective temperature of 3800 K is to the conclusions of \citet{hh14}, which also identifies TW Hya as an M0.5 spectral type.  Also in agreement, the best matching single temperature template to the HST STIS spectrum of TW Hya by \citet{debes13} was to an M0V star with T$_{eff} =$ 3730 K -- this converts to 3900 K with the same conversion from \citet{hh14}, and both effective temperature estimates are within our uncertainties. Unfortunately, neither  \citet{debes13} nor  \citet{hh14} reported a measurement of the surface gravity, however the surface gravity found by \citet{ab02} from initial fits of optical photospheric lines was $\log {\rm g} =4.44 \pm 0.05$, in reasonable agreement with our best-fit value.

Therefore, we put our findings into context by comparing to the two extremes: a study using optical data as well as the near-infrared work of VS11.   \citet{yjv05} (YJV05) found the effective temperature and surface gravity by fitting high resolution optical spectra, finding $T_{\rm eff} = 4126 $ K and $\log {\rm g} =4.8$, and the magnetic field by CO and Ti lines observed in a moderate resolution infrared spectrum, finding the mean magnetic field to be B$=2.7$kG.  This corresponds to a spectral type of K6.5 by YJV05, or K6 using the conversion from \citet{hh14}.  VS11 use an R$=$2000-2500 near-infrared spectrum to spectral type TW Hya as an M2.5, and then conclude it has $T_{\rm eff} = 3400$ K. Alternatively, the conversion of \citet{hh14} would give $T_{\rm eff} \sim 3500$ K.  After deriving the mass and radius, VS11 determine  $\log {\rm g}  =3.8$. VS11 do not consider the magnetic field in their analysis. Clearly, using the same spectral type to temperature conversion does not change the resulting discrepancies. The derived effective temperature, surface gravity, and magnetic field for YJV05, VS11, and this work can be viewed in Table \ref{table-compare}.
 
 \begin{deluxetable}{llll}[h]
\tabletypesize{\footnotesize}
\tablecolumns{8} 
\tablewidth{0pt}
 \tablecaption{The Stellar Parameters of TW Hya \label{table-compare}}
 \tablehead{
\colhead{Reference} & \colhead{$T_{\rm eff}$} & \colhead{$\log {\rm g} $} & \colhead{B} \\
\colhead{} & \colhead{ [K]} & \colhead{} & \colhead{[kG]} } 

 \startdata 
YJV05 & 4125 & 4.8 & 2.7\\ 
VS11 & 3400 & 3.8 & 0.0\tablenotemark{a} \\
This Work & 3800 & 4.2 & 3.0\\
\enddata

 \tablecomments{The results of select TW Hya studies, representing an optical study and an near-IR study, to which our results can be compared.}
\tablenotetext{a}{Magnetic effects were not considered. A value of 0.0 has been adopted for comparison.}
\end{deluxetable}
 
 We compare the YJV05 and VS11 results to our own by synthesizing spectra with MoogStokes using the stellar parameters from these earlier studies. The Na, $^{12}$CO, and Ti intervals are shown in Figure \ref{fig-blended}. We plot the observed IGRINS spectrum of TW Hya in black and overplot the models corresponding to the parameters of VS11, YJV05, and this work in various colors. A veiling of r$_{k} \sim$ 0.3-0.4 is necessary to fit all three of these models.

\begin{figure}[h!]
\includegraphics*[width=0.5\textwidth,angle=0]{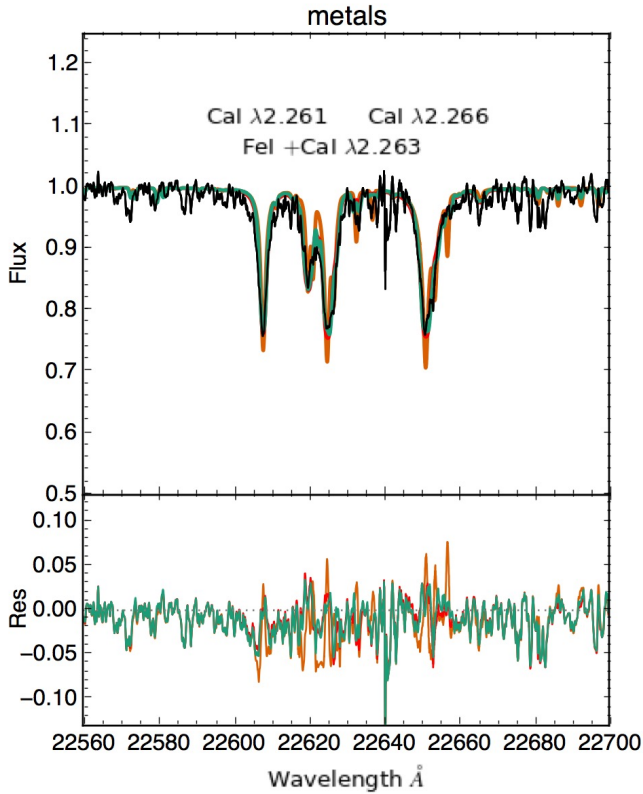}

\caption{\label{fig-blended-metals} Same as Figure \ref{fig-blended}, here showing CaI and FeI lines near the $^{12}$CO interval. This figure shows that the value of the veiling (obtained at 2.2$\mu$m) and our treatment of it produces a quality fit to the observed spectrum to the metal lines near the  $^{12}$CO interval. Thus, we have shown that our treatment of the veiling is adequate across the region we use for our spectral analysis (from the Na interval to the $^{12}$CO interval). As the veiling is found independently, we break the common degeneracy between the effects of veiling and surface gravity when using the $^{12}$CO interval.}
\end{figure}

It is immediately obvious from the plots that our findings ($T_{\rm eff} = 3800 $ K, $\log {\rm g} =4.2$, and B$=3.0$ kG) are in better agreement with YJV05 than with VS11. This is largely due to the inclusion of the magnetic field and the absence of the cool temperature Sc lines near the Na interval. The magnetic field strength determined by YJV05 and this paper are remarkably similar despite our use of different fitting techniques,  different line synthesis programs and stellar atmospheres, and despite the different assumptions about the uniformity of the field. The difference in the determined effective temperatures is $\sim$ 300 K. One could explain that the difference in effective temperatures, with a slightly lower effective temperature observed in the infrared compared to the optical, could be due to stellar spots -- a topic further discussed below. Lastly, the difference in the surface gravity is also likely an improvement on the measurement, as YJV05 note that the value they obtain is rather high, and the residuals from the IGRINS spectrum over the CO interval between the MoogStokes model with the YJV05 parameters are larger than our best-fit model. Another important point here is  that while the effects of veiling and surface gravity are often degenerate in the CO interval, we are able to disentangle our measurements by determining the veiling independently. The best-fit veiling is found at 2.2$\mu$m for each MoogStokes model; then each model is artificially veiled with a blackbody. The validity of this process is demonstrated by the quality of model fits of metal lines near the $^{12}$CO interval in Figure \ref{fig-blended-metals}, which shows that the veiling treatment is satisfactory out to the $^{12}$CO interval wavelength range (and thus over the entire wavelength range of our analysis).

The conclusions of  VS11 are quite different from ours despite the use of  near-infrared spectra by both projects. One contribution to the differences is that VS11 used, in part, the broad band continuum for spectral typing, which can be altered by the combination of different physical effects (effective temperature, veiling, etc.), and therefore is less sensitive to individual parameters. Furthermore, it is clear that  VS11 was impacted by having lower spectral resolution (R $\sim$ 2000-2500) and by omitting the magnetic field.  The expected line profiles resulting from the surface gravity and magnetic parameters can be better understood by considering Figure \ref{fig-intervals-diffsens}. The magnetic field will split the cores of the Na lines at 2.206 and 2.208 $\mu$m, as well as alter the shape of the Ti line near 2.296 $\mu$m in the $^{12}$CO bandhead -- both of these effects are present in our observations and best fit model but are missing from that of VS11. Additionally, the  $^{12}$CO interval (see Fig. \ref{fig-blended}) is not well matched by models that use the parameters of VS11. The predicted lines are too deep, an effect mostly due to the surface gravity (albeit with some contribution from the different effective temperature as well). Lastly, the model corresponding to the lower temperature of 3400 K exhibits strong ScI lines on both sides of the Na I line at 2.206 $\mu$m (see Figures \ref{fig-rms-temp} and \ref{fig-blended}) that are not observed in the IGRINS spectrum. 

\begin{figure*}[t!]
\includegraphics*[width=0.95\textwidth,angle=0]{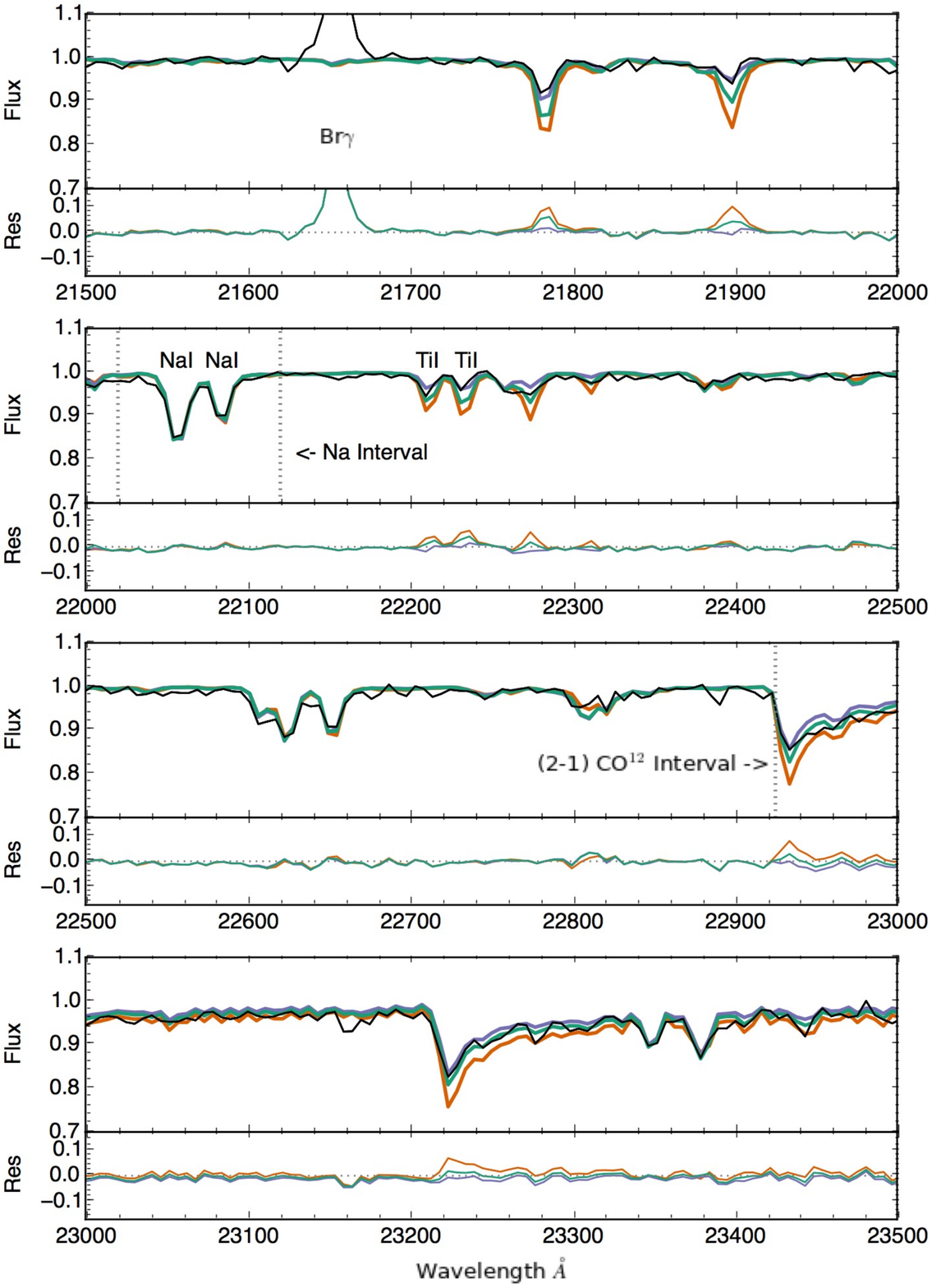}
\caption{\label{fig-irtf} An IRTF spectrum (R $\sim$ 2000-3000) of TW Hya from \citet{covey10} demonstrates the importance of high resolution. We plot matching MoogStokes synthetic spectra (the colors are the same as Figure \ref{fig-blended}) corresponding to the best fit parameters of this work, YJV05, and VS11. The MoogStokes spectra shown here have been convolved to a resolution of R$=$3000 and rebinned to match the IRTF wavelength solution. At this resolution, the model spectra appear identical over the Na interval, which is used in this work to determine the effective temperature, yet clearly show agreement with a magnetic model.}
\end{figure*}

To examine the impact of spectral resolution and the discrepant results of VS11, we found a moderate resolution near-infrared spectrum to compare our and the VS11 parameters. We use an available NASA Infrared Telescope Facility (IRTF) spectrum from \citet{covey10}. This spectrum was obtained in 2008 with the same instrument \citep[SpeX; ][]{rayner03} as that of VS11 at nearly the same spectral resolution of R $\sim$ 2000-3000 (although it is lower quality). We convolve the MoogStokes synthetic spectra corresponding to the findings of YJV05, VS11, and this work to the larger resolution (R$=$3000) and rebin to the observed IRTF wavelength solution. The veiling is similar to the IGRINS work above (r$_{k} \approx$ 0.2-0.4 at 2.2$\mu$m). The resulting comparisons between the observed IRTF spectrum and lower resolution models are quite intriguing. The synthetic spectra are essentially identical over the Na interval at this spectral resolution, and therefore this plot shows that this sensitive region can not be used (as in this work) for finding the effective temperature without adequate spectral resolution. However, the Ti lines observed in the IRTF spectrum clearly do require the inclusion of a magnetic field, and the fit to $^{12}$CO interval is best with our best fit model. Ultimately, this exercise suggests that the spectral analysis can be greatly limited by spectral resolution, and even so, our results are preferable to that of VS11.

One final consideration in comparing  derived spectral types of TW Hya is that a single effective temperature may not be adequate to describe its stellar surface. \citet{debes13} find a somewhat better match to the observed 550-1000 nm spectrum of TW Hya using a 45\%/55\% flux-weighted blend of K7 and M2.5 template star spectra than they find when matching with any single template spectrum. VS11 interpreted TW Hya as a cool star with a hot accretion spot. Alternatively, TW Hya also has been listed as a likely comparison to heavily-spotted stars \citep{gully17}. For example, \citet{gully17} reviews an IGRINS spectrum of LkCa 14 and finds a 4100 K hot photosphere component mixed with a 2700-3000 K cool component likely from stellar spots. However, LkCa 14 is unusual in that the stellar spot covering fraction of 80\% is derived, producing a temperature contrast that is more readily detected.  Variability would be expected in cool spots or hot accretion spots.  \citet{hue08} suggest that observed RV modulations in TW Hya may be due to stellar spots rather than an exoplanet \citep{hue08}. Thus, perhaps a comparison to LkCa 14 is reasonable, except that any spotting would be much less extreme: \citet{hue08} finds that only 7\% of the surface may be covered by a cold spot. While we fit the IGRINS near-IR spectrum with a single temperature model in this work, the presence of spots, either hot or cool, is very probable and a two temperature scheme may be preferable if higher signal-to-noise data are obtained in the future.

\subsection{Age of TW Hya}

\begin{figure}[t!]
\includegraphics*[width=0.5\textwidth,angle=0]{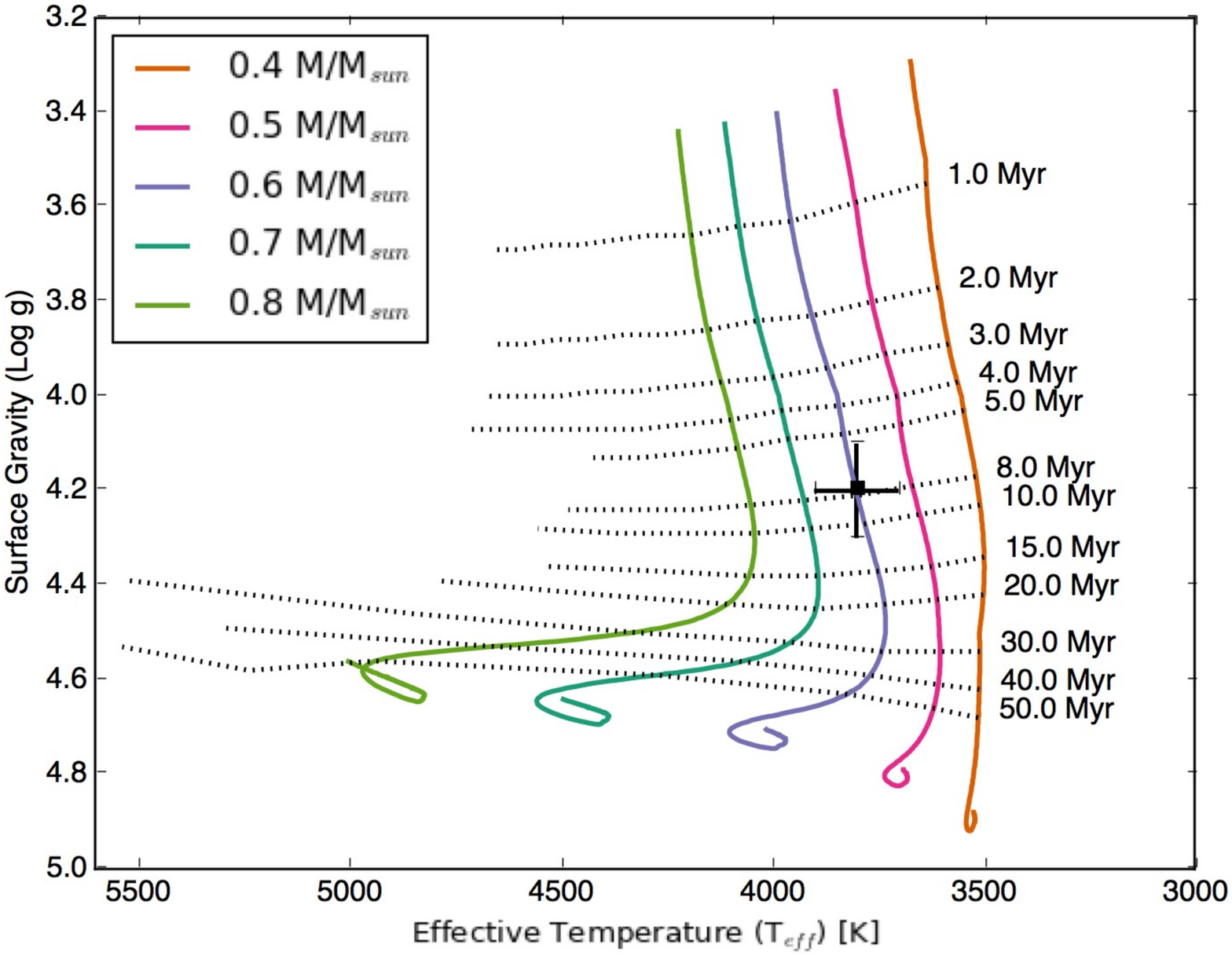}
\caption{\label{fig-hr} The location of TW Hya on the HR diagram using the best fit parameters of the IGRINS spectrum. The \citet{bar15} evolutionary tracks for stars from 0.4 - 0.8 M$_\sun$ are plotted in color and isochrones from 1 to 50 Myr are plotted with dashed lines and labeled.  }
\end{figure}

We evaluate the fundamental characteristics of TW Hya by plotting our best fit parameters on the spectroscopist's H-R diagram in Figure \ref{fig-hr}. We use the evolutionary tracks and isochrones of \citet{bar15}. From the location of TW Hya on this H-R diagram, we find TW Hya is a $0.6 \pm 0.1$ M$_\sun$ star and has an age of $8 \pm 3$  Myr. 

Use of TW Hya's measured flux also agrees with these findings. If we correct the bolometric luminosity values derived by VS11 from 2MASS J-band photometry and by \citet{hh14} from their own flux measurements to the new distance of 59.5pc  \citep{gaia_b}, the luminosity of TW Hya is L$_{\star}$ $=$ 0.23 L$_{\sun}$. YJV05 did not explain their adopted  luminosity, and we therefore exclude it. Using the same evolutionary tracks plotted with luminosity versus effective temperature or surface gravity, then TW Hya is consistent with a $0.6 \pm 0.1$ M$_\sun$ and $6 \pm 3$  Myr old star and a $0.65 \pm 0.15$ M$_\sun$ and $8 \pm 3$  Myr old star, respectively. Both of these results are in agreement with our measurements using the spectroscopist's HR diagram. Lastly, the observed photometry also does not rule out our measured veiling component, as the range of the predicted K magnitude values from the  same evolutionary tracks allows for the contribution from blackbody emission at 1500 K.

Overall, our derivation of the mass and age of TW Hya agree reasonably well with the literature. Masses available in the literature span from 0.4 - 0.9 M$_\sun$ (e.g. VS11 and YJV05). Our derived age of 8 Myr clearly fits in the middle of other reported results, from 3 Myr (VS11) to 16 Myr \citep{hh14}. To be somewhat consistent with our new result, if we place the $T_{\rm eff}$ and $\log {\rm g} $ values derived by YJV05 and VS11 on the new \citet{bar15} isochrones, we derive ages of $>50$ Myr (YJV05 note that $\log {\rm g}$ is high) and $\sim$ 2 Myr, respectively. 

Specifically, we see very good agreement with a couple of studies. However, it should be mentioned that many improvements, largely regarding new evolutionary tracks and isochrones from \citet{bar15} and a more accurate distance from \citet{gaia_b}, have been made since these previous works were published. Regardless, \citet{debes13} infer equivalent results to ours, that TW Hya has a stellar mass of 0.55 M$_\sun$ and age of 8 Myr. This agreement with \citet{debes13} comes with a caveat due to differences in method: the radius is from a comparison star (used to plot on the H-R diagram rather than surface gravity), a dominant cool temperature star of $T_{\rm eff} \sim$ 3600K is assumed, and the older \citet{bar98} models were used. Another study by \citet{wein13} estimate the age of TW Hya to be $6 \pm 3$ Myr, to which our results agrees within the uncertainties as well. Additionally, our determined age of 8 Myr for TW Hya is also in excellent agreement with the median age of the TW Hydra cluster, as determined by parallaxes and kinematics \citep[7.5 Myr and 7.9 Myr; ][]{duc14, don16}. 

\section{Conclusions}

We have presented a high resolution near-IR spectrum of TW Hya obtained with IGRINS  at the 2.7m Harlan J. Smith Telescope at McDonald Observatory over several epochs. We compare these high quality data to synthetic spectra from MoogStokes, which includes magnetic effects while computing the emergent stellar spectra. After identifying spectral regions sensitive to changes in specific stellar parameters, we find the best fit MoogStokes synthetic spectrum corresponds to the stellar parameters of  $T_{\rm eff} = 3800 \pm100$ K, $\log \rm{g}=4.2 \pm 0.1$, and B$=3.0 \pm 0.2$ kG. Our parameterization of TW Hya does not rely on spectral type or distance measurements. This work confirms that TW Hya has a spectral type of $\sim$ M0.5 in agreement with \citet{hh14} and resolves the debate between the optical spectral type of K6.5-7 (e.g. YJV05) and the sole conflicting near-IR derived type of M2.5 (VS11).  Adoption of our derived stellar parameters leads to a mass of $0.6 \pm 0.1$ M$_\sun$ and an age of $8  \pm 3$ Myr, independent from distance estimates. 

Almost all analysis of TW Hya, and much of that of the disk as well, relies on an assumption of some stellar parameter that we have now measured using the high resolution near-IR spectrum. For instance, a stellar mass is necessary to calculate the photoionization impacting the disk \citep{erc17}. More significantly, \citet{dnc17} input effective temperatures and surface gravities to identify Moog models for M-band spectra to search for CO v=1-0 emission (and thus gas in the disk). The presence of residual gas would have implications for the disk dispersal timescales and the role of the gas in planet formation. These two quick examples illustrate that the stellar parameters are a critical element to not only defining the properties in a system, but for investigating the physical state. Ultimately, understanding how the disks of HL Tau and TW Hya exhibit such similar structure, while TW Hya is 2-3 times the median age of gaseous disks, requires a better understanding of the stars themselves. As ALMA and other millimeter observations expand our knowledge of stellar disks, it will be evermore important for additional investigations into the host stars (similar to this work regarding TW Hya) to interpret and deepen our understanding the process of star and disk formation and evolution.

\acknowledgements

We thank the anonymous referee for their insightful comments. This work used the Immersion Grating Infrared Spectrograph (IGRINS) that was developed under a collaboration between the University of Texas at Austin and the Korea Astronomy and Space Science Institute (KASI) with the financial support of the US National Science Foundation under grant AST-1229522, of the University of Texas at Austin, and of the Korean GMT Project of KASI.

This work has made use of data from the European Space Agency (ESA)
mission {\it Gaia} (\url{https://www.cosmos.esa.int/gaia}), processed by
the {\it Gaia} Data Processing and Analysis Consortium (DPAC,
\url{https://www.cosmos.esa.int/web/gaia/dpac/consortium}). Funding
for the DPAC has been provided by national institutions, in particular
the institutions participating in the {\it Gaia} Multilateral Agreement.

\software{IGRINS pipeline package (version 2.1 alpha 3; Lee \& Gullikson 2016), MoogStokes (Deen 2013), MARCS (Gustafsson et al. 2008)}
\facility{Smith (IGRINS) -- McDonald Observatory's 2.7m Harlan J. Smith Telescope}

\bibliography{twhya}

\begin{thebibliography}{}
\expandafter\ifx\csname natexlab\endcsname\relax\def\natexlab#1{#1}\fi
\providecommand{\url}[1]{\href{#1}{#1}}

\bibitem[{{Alencar} \& {Batalha}(2002)}]{ab02}
{Alencar}, S.~H.~P., \& {Batalha}, C. 2002, \apj, 571, 378

\bibitem[{{ALMA Partnership} {et~al.}(2015){ALMA Partnership}, {Brogan},
  {P{\'e}rez}, {Hunter}, {Dent}, {Hales}, {Hills}, {Corder}, {Fomalont},
  {Vlahakis}, {Asaki}, {Barkats}, {Hirota}, {Hodge}, {Impellizzeri}, {Kneissl},
  {Liuzzo}, {Lucas}, {Marcelino}, {Matsushita}, {Nakanishi}, {Phillips},
  {Richards}, {Toledo}, {Aladro}, {Broguiere}, {Cortes}, {Cortes}, {Espada},
  {Galarza}, {Garcia-Appadoo}, {Guzman-Ramirez}, {Humphreys}, {Jung}, {Kameno},
  {Laing}, {Leon}, {Marconi}, {Mignano}, {Nikolic}, {Nyman}, {Radiszcz},
  {Remijan}, {Rod{\'o}n}, {Sawada}, {Takahashi}, {Tilanus}, {Vila Vilaro},
  {Watson}, {Wiklind}, {Akiyama}, {Chapillon}, {de Gregorio-Monsalvo}, {Di
  Francesco}, {Gueth}, {Kawamura}, {Lee}, {Nguyen Luong}, {Mangum}, {Pietu},
  {Sanhueza}, {Saigo}, {Takakuwa}, {Ubach}, {van Kempen}, {Wootten},
  {Castro-Carrizo}, {Francke}, {Gallardo}, {Garcia}, {Gonzalez}, {Hill},
  {Kaminski}, {Kurono}, {Liu}, {Lopez}, {Morales}, {Plarre}, {Schieven},
  {Testi}, {Videla}, {Villard}, {Andreani}, {Hibbard}, \&
  {Tatematsu}}]{brogan15}
{ALMA Partnership}, {Brogan}, C.~L., {P{\'e}rez}, L.~M., {et~al.} 2015, \apjl,
  808, L3

\bibitem[{{Andrews} {et~al.}(2016){Andrews}, {Wilner}, {Zhu}, {Birnstiel},
  {Carpenter}, {P{\'e}rez}, {Bai}, {{\"O}berg}, {Hughes}, {Isella}, \&
  {Ricci}}]{awz15}
{Andrews}, S.~M., {Wilner}, D.~J., {Zhu}, Z., {et~al.} 2016, \apjl, 820, L40

\bibitem[{{Baraffe} {et~al.}(1998){Baraffe}, {Chabrier}, {Allard}, \&
  {Hauschildt}}]{bar98}
{Baraffe}, I., {Chabrier}, G., {Allard}, F., \& {Hauschildt}, P.~H. 1998, \aap,
  337, 403

\bibitem[{{Baraffe} {et~al.}(2015){Baraffe}, {Homeier}, {Allard}, \&
  {Chabrier}}]{bar15}
{Baraffe}, I., {Homeier}, D., {Allard}, F., \& {Chabrier}, G. 2015, \aap, 577,
  A42

\bibitem[{{Bergin} {et~al.}(2013){Bergin}, {Cleeves}, {Gorti}, {Zhang},
  {Blake}, {Green}, {Andrews}, {Evans}, {Henning}, {{\"O}berg}, {Pontoppidan},
  {Qi}, {Salyk}, \& {van Dishoeck}}]{ber13}
{Bergin}, E.~A., {Cleeves}, L.~I., {Gorti}, U., {et~al.} 2013, \nat, 493, 644

\bibitem[{{Brice{\~n}o} {et~al.}(2002){Brice{\~n}o}, {Luhman}, {Hartmann},
  {Stauffer}, \& {Kirkpatrick}}]{bri02}
{Brice{\~n}o}, C., {Luhman}, K.~L., {Hartmann}, L., {Stauffer}, J.~R., \&
  {Kirkpatrick}, J.~D. 2002, \apj, 580, 317

\bibitem[{{Cieza} {et~al.}(2005){Cieza}, {Kessler-Silacci}, {Jaffe}, {Harvey},
  \& {Evans}}]{cieza05}
{Cieza}, L.~A., {Kessler-Silacci}, J.~E., {Jaffe}, D.~T., {Harvey}, P.~M., \&
  {Evans}, II, N.~J. 2005, \apj, 635, 422

\bibitem[{{Covey} {et~al.}(2010){Covey}, {Lada}, {Rom{\'a}n-Z{\'u}{\~n}iga},
  {Muench}, {Forbrich}, \& {Ascenso}}]{covey10}
{Covey}, K.~R., {Lada}, C.~J., {Rom{\'a}n-Z{\'u}{\~n}iga}, C., {et~al.} 2010,
  \apj, 722, 971

\bibitem[{{Cushing} {et~al.}(2008){Cushing}, {Marley}, {Saumon}, {Kelly},
  {Vacca}, {Rayner}, {Freedman}, {Lodders}, \& {Roellig}}]{cush08}
{Cushing}, M.~C., {Marley}, M.~S., {Saumon}, D., {et~al.} 2008, \apj, 678, 1372

\bibitem[{{Debes} {et~al.}(2013){Debes}, {Jang-Condell}, {Weinberger},
  {Roberge}, \& {Schneider}}]{debes13}
{Debes}, J.~H., {Jang-Condell}, H., {Weinberger}, A.~J., {Roberge}, A., \&
  {Schneider}, G. 2013, \apj, 771, 45

\bibitem[{{Deen}(2013)}]{deen13}
{Deen}, C.~P. 2013, \aj, 146, 51

\bibitem[{{Donaldson} {et~al.}(2016){Donaldson}, {Weinberger}, {Gagn{\'e}},
  {Faherty}, {Boss}, \& {Keiser}}]{don16}
{Donaldson}, J.~K., {Weinberger}, A.~J., {Gagn{\'e}}, J., {et~al.} 2016, \apj,
  833, 95

\bibitem[{{Doppmann} \& {Jaffe}(2003)}]{dj03}
{Doppmann}, G.~W., \& {Jaffe}, D.~T. 2003, \aj, 126, 3030

\bibitem[{{Doppmann} {et~al.}(2017){Doppmann}, {Najita}, \& {Carr}}]{dnc17}
{Doppmann}, G.~W., {Najita}, J.~R., \& {Carr}, J.~S. 2017, \apj, 836, 242

\bibitem[{{Ducourant} {et~al.}(2014){Ducourant}, {Teixeira}, {Galli}, {Le
  Campion}, {Krone-Martins}, {Zuckerman}, {Chauvin}, \& {Song}}]{duc14}
{Ducourant}, C., {Teixeira}, R., {Galli}, P.~A.~B., {et~al.} 2014, \aap, 563,
  A121

\bibitem[{{Endl} \& {Cochran}(2016)}]{ec16}
{Endl}, M., \& {Cochran}, W.~D. 2016, \pasp, 128, 094502

\bibitem[{{Ercolano} {et~al.}(2017){Ercolano}, {Rosotti}, {Picogna}, \&
  {Testi}}]{erc17}
{Ercolano}, B., {Rosotti}, G.~P., {Picogna}, G., \& {Testi}, L. 2017, \mnras,
  464, L95

\bibitem[{{Evans} {et~al.}(2009){Evans}, {Dunham}, {J{\o}rgensen}, {Enoch},
  {Mer{\'{\i}}n}, {van Dishoeck}, {Alcal{\'a}}, {Myers}, {Stapelfeldt},
  {Huard}, {Allen}, {Harvey}, {van Kempen}, {Blake}, {Koerner}, {Mundy},
  {Padgett}, \& {Sargent}}]{ev09}
{Evans}, II, N.~J., {Dunham}, M.~M., {J{\o}rgensen}, J.~K., {et~al.} 2009,
  \apjs, 181, 321

\bibitem[{{Gaia Collaboration} {et~al.}(2016{\natexlab{a}}){Gaia
  Collaboration}, {Prusti}, {de Bruijne}, {Brown}, {Vallenari}, {Babusiaux},
  {Bailer-Jones}, {Bastian}, {Biermann}, {Evans}, \& et~al.}]{gaia_a}
{Gaia Collaboration}, {Prusti}, T., {de Bruijne}, J.~H.~J., {et~al.}
  2016{\natexlab{a}}, \aap, 595, A1

\bibitem[{{Gaia Collaboration} {et~al.}(2016{\natexlab{b}}){Gaia
  Collaboration}, {Brown}, {Vallenari}, {Prusti}, {de Bruijne}, {Mignard},
  {Drimmel}, {Babusiaux}, {Bailer-Jones}, {Bastian}, \& et~al.}]{gaia_b}
{Gaia Collaboration}, {Brown}, A.~G.~A., {Vallenari}, A., {et~al.}
  2016{\natexlab{b}}, \aap, 595, A2

\bibitem[{{Gully-Santiago} {et~al.}(2017){Gully-Santiago}, {Herczeg},
  {Czekala}, {Somers}, {Grankin}, {Covey}, {Donati}, {Alencar}, {Hussain},
  {Shappee}, {Mace}, {Lee}, {Holoien}, {Jose}, \& {Liu}}]{gully17}
{Gully-Santiago}, M.~A., {Herczeg}, G.~J., {Czekala}, I., {et~al.} 2017, ArXiv
  e-prints, arXiv:1701.06703

\bibitem[{{Gustafsson} {et~al.}(2008){Gustafsson}, {Edvardsson}, {Eriksson},
  {J{\o}rgensen}, {Nordlund}, \& {Plez}}]{gus08}
{Gustafsson}, B., {Edvardsson}, B., {Eriksson}, K., {et~al.} 2008, \aap, 486,
  951

\bibitem[{{Herbig}(1978)}]{herbig78}
{Herbig}, G.~H. 1978, {Can Post-T Tauri Stars Be Found?}, ed. L.~V. {Mirzoyan},
  171

\bibitem[{{Herczeg} \& {Hillenbrand}(2014)}]{hh14}
{Herczeg}, G.~J., \& {Hillenbrand}, L.~A. 2014, \apj, 786, 97

\bibitem[{{Hu{\'e}lamo} {et~al.}(2008){Hu{\'e}lamo}, {Figueira}, {Bonfils},
  {Santos}, {Pepe}, {Gillon}, {Azevedo}, {Barman}, {Fern{\'a}ndez}, {di Folco},
  {Guenther}, {Lovis}, {Melo}, {Queloz}, \& {Udry}}]{hue08}
{Hu{\'e}lamo}, N., {Figueira}, P., {Bonfils}, X., {et~al.} 2008, \aap, 489, L9

\bibitem[{{Johns-Krull} {et~al.}(2009){Johns-Krull}, {Greene}, {Doppmann}, \&
  {Covey}}]{jk09}
{Johns-Krull}, C.~M., {Greene}, T.~P., {Doppmann}, G.~W., \& {Covey}, K.~R.
  2009, \apj, 700, 1440

\bibitem[{Lee \& Gullikson(2016)}]{lee15}
Lee, J.-J., \& Gullikson, K. 2016, plp: v2.1 alpha 3, ,

\bibitem[{{Mace} {et~al.}(2016){Mace}, {Kim}, {Jaffe}, {Park}, {Lee}, {Kaplan},
  {Yu}, {Yuk}, {Chun}, {Pak}, {Kim}, {Lee}, {Sneden}, {Afsar}, {Pavel}, {Lee},
  {Oh}, {Jeong}, {Park}, {Kidder}, {Lee}, {Nguyen Le}, {McLane},
  {Gully-Santiago}, {Oh}, {Lee}, {Hwang}, \& {Park}}]{mace16}
{Mace}, G., {Kim}, H., {Jaffe}, D.~T., {et~al.} 2016, in \procspie, Vol. 9908,
  Society of Photo-Optical Instrumentation Engineers (SPIE) Conference Series,
  99080C

\bibitem[{{Malo} {et~al.}(2014){Malo}, {Doyon}, {Feiden}, {Albert},
  {Lafreni{\`e}re}, {Artigau}, {Gagn{\'e}}, \& {Riedel}}]{malo14}
{Malo}, L., {Doyon}, R., {Feiden}, G.~A., {et~al.} 2014, \apj, 792, 37

\bibitem[{{Manara} {et~al.}(2017){Manara}, {Frasca}, {Alcala}, {Natta},
  {Stelzer}, \& {Testi}}]{man17}
{Manara}, C.~F., {Frasca}, A., {Alcala}, J.~M., {et~al.} 2017, ArXiv e-prints,
  arXiv:1705.10075

\bibitem[{{Mohanty} {et~al.}(2004){Mohanty}, {Basri}, {Jayawardhana}, {Allard},
  {Hauschildt}, \& {Ardila}}]{mohanty04}
{Mohanty}, S., {Basri}, G., {Jayawardhana}, R., {et~al.} 2004, \apj, 609, 854

\bibitem[{{Padgett}(1996)}]{padgett96}
{Padgett}, D.~L. 1996, \apj, 471, 847

\bibitem[{{Park} {et~al.}(2014){Park}, {Jaffe}, {Yuk}, {Chun}, {Pak}, {Kim},
  {Pavel}, {Lee}, {Oh}, {Jeong}, {Sim}, {Lee}, {Nguyen Le}, {Strubhar},
  {Gully-Santiago}, {Oh}, {Cha}, {Moon}, {Park}, {Brooks}, {Ko}, {Han}, {Nah},
  {Hill}, {Lee}, {Barnes}, {Yu}, {Kaplan}, {Mace}, {Kim}, {Lee}, {Hwang}, \&
  {Park}}]{park14}
{Park}, C., {Jaffe}, D.~T., {Yuk}, I.-S., {et~al.} 2014, in Society of
  Photo-Optical Instrumentation Engineers (SPIE) Conference Series, Vol. 9147,
  Society of Photo-Optical Instrumentation Engineers (SPIE) Conference Series,
  1

\bibitem[{{Rayner} {et~al.}(2003){Rayner}, {Toomey}, {Onaka}, {Denault},
  {Stahlberger}, {Vacca}, {Cushing}, \& {Wang}}]{rayner03}
{Rayner}, J.~T., {Toomey}, D.~W., {Onaka}, P.~M., {et~al.} 2003, \pasp, 115,
  362

\bibitem[{{Rice} {et~al.}(2010){Rice}, {Barman}, {Mclean}, {Prato}, \&
  {Kirkpatrick}}]{rice10}
{Rice}, E.~L., {Barman}, T., {Mclean}, I.~S., {Prato}, L., \& {Kirkpatrick},
  J.~D. 2010, \apjs, 186, 63

\bibitem[{{Santos} {et~al.}(2008){Santos}, {Melo}, {James}, {Gameiro},
  {Bouvier}, \& {Gomes}}]{santos08}
{Santos}, N.~C., {Melo}, C., {James}, D.~J., {et~al.} 2008, \aap, 480, 889

\bibitem[{{Sneden}(1973)}]{sne73}
{Sneden}, C.~A. 1973, PhD thesis, THE UNIVERSITY OF TEXAS AT AUSTIN.

\bibitem[{{Stelzer} {et~al.}(2013){Stelzer}, {Frasca}, {Alcal{\'a}}, {Manara},
  {Biazzo}, {Covino}, {Rigliaco}, {Testi}, {Covino}, \& {D'Elia}}]{stelzer13}
{Stelzer}, B., {Frasca}, A., {Alcal{\'a}}, J.~M., {et~al.} 2013, \aap, 558,
  A141

\bibitem[{{Tsukagoshi} {et~al.}(2016){Tsukagoshi}, {Nomura}, {Muto}, {Kawabe},
  {Ishimoto}, {Kanagawa}, {Okuzumi}, {Ida}, {Walsh}, \& {Millar}}]{tsu16}
{Tsukagoshi}, T., {Nomura}, H., {Muto}, T., {et~al.} 2016, \apjl, 829, L35

\bibitem[{{Vacca} \& {Sandell}(2011)}]{vs11}
{Vacca}, W.~D., \& {Sandell}, G. 2011, \apj, 732, 8

\bibitem[{{van Boekel} {et~al.}(2017){van Boekel}, {Henning}, {Menu}, {de
  Boer}, {Langlois}, {M{\"u}ller}, {Avenhaus}, {Boccaletti}, {Schmid},
  {Thalmann}, {Benisty}, {Dominik}, {Ginski}, {Girard}, {Gisler}, {Lobo Gomes},
  {Menard}, {Min}, {Pavlov}, {Pohl}, {Quanz}, {Rabou}, {Roelfsema}, {Sauvage},
  {Teague}, {Wildi}, \& {Zurlo}}]{vb17}
{van Boekel}, R., {Henning}, T., {Menu}, J., {et~al.} 2017, \apj, 837, 132

\bibitem[{{Webb} {et~al.}(1999){Webb}, {Zuckerman}, {Platais}, {Patience},
  {White}, {Schwartz}, \& {McCarthy}}]{webb99}
{Webb}, R.~A., {Zuckerman}, B., {Platais}, I., {et~al.} 1999, \apjl, 512, L63

\bibitem[{{Weinberger} {et~al.}(2013){Weinberger}, {Anglada-Escud{\'e}}, \&
  {Boss}}]{wein13}
{Weinberger}, A.~J., {Anglada-Escud{\'e}}, G., \& {Boss}, A.~P. 2013, \apj,
  762, 118

\bibitem[{{Williams} \& {Cieza}(2011)}]{wc11}
{Williams}, J.~P., \& {Cieza}, L.~A. 2011, \araa, 49, 67

\bibitem[{{Yang} {et~al.}(2005){Yang}, {Johns-Krull}, \& {Valenti}}]{yjv05}
{Yang}, H., {Johns-Krull}, C.~M., \& {Valenti}, J.~A. 2005, \apj, 635, 466

\end{thebibliography}

\end{document}